\documentclass[twocolumn,
nofootinbib,superscriptaddress,aps,10pt,longbibliography]{revtex4-1}

\usepackage{bbold}

\usepackage[usenames,dvipsnames]{xcolor}
\usepackage{amsmath}
\usepackage{amsthm, amssymb}
\usepackage{enumitem}
\usepackage{slashed}
\usepackage{graphicx}
\usepackage{tikz}
\usetikzlibrary{calc,fadings,decorations.pathreplacing,shapes,shapes.multipart,arrows,shapes.misc,intersections,positioning,patterns}
\usepackage{bm}
\usepackage{dsfont}
\usepackage{changepage}
\usepackage{array}
\usepackage{appendix}

\definecolor{bluepurple2}{rgb}{0.06,0,0.6}
\usepackage[colorlinks=true,citecolor=blue,linkcolor=bluepurple2]{hyperref}

\usepackage{bm}
\renewcommand{\vec}[1]{\boldsymbol{\mathbf{#1}}}

\graphicspath{{./}{./images/}}

\newcommand{\bit}{\begin{itemize}}
\newcommand{\eit}{\end{itemize}}

\newcommand{\f}{\frac}
\renewcommand{\>}{\right\rangle}
\newcommand{\<}{\left\langle}
\newcommand{\ba}{\begin{align}}
\newcommand{\ea}{\end{align}}
\newcommand{\be}{\begin{equation}}
\newcommand{\ee}{\end{equation}}
\newcommand{\bi}{\begin{itemize}}
\newcommand{\ei}{\end{itemize}}
\newcommand{\lf}{\left(}
\newcommand{\ri}{\right)}
\newcommand{\dd}{\mathrm{d}}

\newcommand{\Tr}{\operatorname{Tr}}

\newcommand{\blue}{\color{blue}}

\begin{document}
\date{\today}

\newcommand{\bra}[1]{\< #1 \right|}
\newcommand{\ket}[1]{\left| #1 \>}

\newcommand{\comment}[1]{{\blue [#1]}}

\title{Spacetime picture for entanglement generation in noisy fermion chains}

\author{Tobias Swann}
\affiliation{Rudolf Peierls Centre for Theoretical Physics, Clarendon Laboratory, Parks Road,
Oxford OX1 3PU, UK}

\author{Denis Bernard}
\affiliation{Laboratoire de Physique de l’\'Ecole Normale Sup\'erieure, CNRS, ENS \& Universit\'e PSL, Sorbonne Universit\'e, Universit\'e Paris Cit\'e, 75005 Paris, France}

\author{Adam Nahum}
\affiliation{Laboratoire de Physique de l’\'Ecole Normale Sup\'erieure, CNRS, ENS \& Universit\'e PSL, Sorbonne Universit\'e, Universit\'e Paris Cit\'e, 75005 Paris, France}
\affiliation{Rudolf Peierls Centre for Theoretical Physics, Clarendon Laboratory, Parks Road,
Oxford OX1 3PU, UK}

\date{\today}

\begin{abstract}
Studies of random unitary circuits have shown that the calculation of R\'enyi entropies of entanglement can be mapped to classical statistical mechanics problems in spacetime. In this paper, we develop an analogous spacetime picture of entanglement generation for random free or weakly interacting fermion systems without conservation laws. We first study a free-fermion model, namely a 1D chain of Majorana modes with nearest neighbour hoppings, random in both space and time. We analyze the $N$th R\'enyi entropy of entanglement using a replica formalism, and we show that the effective model is equivalent to an ${\rm SO}(2N)$ Heisenberg spin chain evolving in imaginary time. By applying a saddle-point approximation to the coherent states path integral for the $N=2$ case, we arrive at a semiclassical picture for the dynamics of the entanglement purity, in terms of two classical fields in spacetime. The classical solutions involve a smooth domain wall that interpolates between two values, with this domain wall relaxing diffusively in the time direction. 
We then study how adding weak interactions to the free-fermion model modifies this spacetime picture, 
reflecting a crossover from diffusive to ballistic spreading of information.
 \end{abstract}
 \maketitle

\section{Introduction}

The aim of this paper will be to develop a spacetime picture for entanglement generation \cite{calabrese2005evolution,islam2015measuring} in random free-fermion systems,
analogous to the scaling picture  in terms of an entanglement ``membrane'' in spacetime that holds for various random and non-random interacting systems \cite{jonay2018coarse, mezei2018membrane, Zhou_2020}, and also to 
explore the crossover between the two ``universality classes'' when weak interactions are switched on.
We find that in the free case, the key spacetime structures are smooth field configurations in a simple effective field theory,
in contrast to the membrane that is relevant in the interacting case
(which in 1+1 dimensions is a spacetime trajectory). 
Turning on interactions leads to field configurations with localized domain wall structures that may be identified with the entanglement membrane.

We will focus on free-fermion systems which are otherwise completely random, meaning there are no additional symmetries which would give rise to conserved charges.
To study the crossover in entanglement dynamics from the non-interacting case to the interacting case, we add weak random interactions to the free-fermion model and see how the effective theory is modified.

Our approach will be to calculate the $N$th R\'enyi entropy of entanglement using a replica formalism, with $N$ forward and $N$ backward Feynman trajectories. 
In this replica formalism, 
an effective model is obtained by averaging over randomness.
For generic   systems, this introduces a  symmetry under discrete permutations of the replicas,
and there is a corresponding effective degree of freedom that takes discrete values
\cite{Nahum_2018,von_Keyserlingk_2018,ZhouNahum,Chan_2018,hunterjones,Zhou_2020,jian2020measurement,bao2020theory,hayden2016holographic,vasseur2019entanglement}. 
By contrast for free fermions there is a larger symmetry under continuous rotations in replica space, which in the most generic case 
is  ${\rm SO}(2N)$. 
This leads to a qualitatively different spacetime picture, with domain walls enforced by the boundary conditions relaxing continuously in the time direction.

To develop the spacetime picture for free fermions, we will study a specific, simple model, which is a one-dimensional chain of Majorana modes with random nearest-neighbour couplings. 
There is only one possible form of such couplings, proportional to product of the two Majorana operators. The coupling strength 
(hopping amplitude)
on each bond is simply a real value, which we take to be random in both space and time.
(We consider the limit where the evolution is in continuous time.)
This model can be thought of as a noisy version of the Kitaev chain \cite{AYuKitaev_2001}
(or equivalently a noisy version of the transverse field Ising model). 
A discrete-time version of this model has previously been studied in Ref.~\cite{bao2021symmetry}, in which the authors  identified the key ${\rm SO}(2N)$ symmetry as well as the ${N=2}$ effective Hamiltonian, 
although there the focus was on the effect of measurements (see Refs.~\cite{bernard2018transport,cao2019entanglement,nahum2020entanglement,alberton2021entanglement,chen2020emergent,nahum2021measurement,zhang2021emergent,sigmamodelmeasurement} for free fermions with measurement).

Under the nickname of Q-SSEP, charge-conserving noisy Hamiltonians have been addressed using a formalism based on a stochastic formulation in \cite{Bauer2017Stochastic,Bauer2019Equilibrium}. There, the dynamics of the $N$ replica theory admits an ${\rm SU
}(2N)$ symmetry \cite{bernard2022dynamics}. Because of the local charge conservation, these model systems allow for transport and can be driven out of equilibrium \cite{Bernard2019Open,bernard2021solution,hruza2022dynamics}.

The model we study however is a model of noisy free fermions with no local conserved quantities. This is the simplest setting: since no conserved local densities exist, the only  fields that will appear in the continuum theory are those encoding entanglement. The only conserved quantity in our model is global fermion parity, which is conserved in any local fermionic model.

After applying the replica formalism to this Majorana chain, we find that the resulting model is equivalent to an ${\rm SO}(2N)$ ferromagnetic Heisenberg chain evolving in imaginary time. When calculating the purity of a section of the Majorana chain ($N=2$), the resulting ${\rm SO}(4)$ Heisenberg chain can be reduced to an ordinary ${\rm SU}(2)$ chain, giving an intuitive picture of an (initially sharp) domain wall relaxing under imaginary time evolution.

This also allows us to reformulate the problem as a spin coherent state path integral in terms of a complex field $z(x,\tau)$. By applying a saddle-point approximation, $z(x,\tau)$ essentially becomes a classical field, giving us a spacetime picture of entanglement generation, in which the entanglement is identified with the classical action of a smooth ``domain wall'' configuration in spacetime.

\begin{figure}
\centering
\includegraphics[width=\columnwidth]{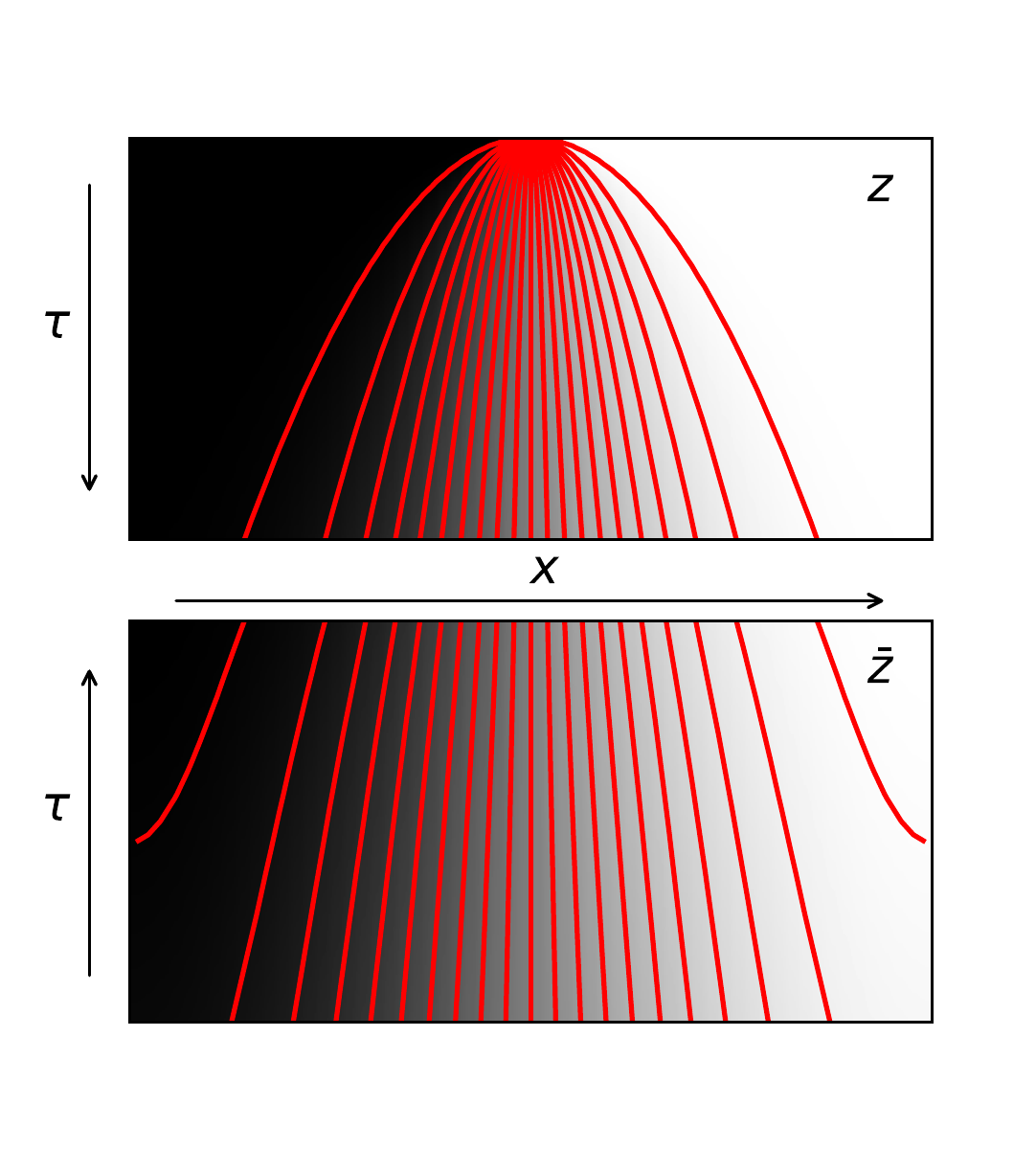}
\caption{ Contour plot of the two independent fields $z$ and $\bar{z}$ involved in calculating the purity of rightmost half of the Majorana chain after time $t$
using a saddle point approximation of the coherent states path integral.
The coordinate $x$ is the physical spatial coordinate.
$\tau$ is a \textit{reversed} time coordinate: i.e. $\tau=0$ at the final  time $t$, and $\tau=t$ at physical time 0 when the system is prepared in a state of short range entanglement.
The field $z$ begins as a sharp domain wall at $\tau=0$ but relaxes diffusively as $\tau$ increases. The field $\bar{z}$ matches $z$ at $\tau=t$, but relaxes diffusively for \emph{decreasing} $\tau$. There are also interactions between the two fields.}\label{fig:contourplot}
\end{figure}

However, an interesting feature of this picture, 
not anticipated from the interacting case, is that the field $z(x,t)$ and its  conjugate field $\bar{z}(x,t)$ must be treated independently. 
Both fields exhibit a kind of diffusive relaxation, but the
field $z$ relaxes in one time direction, while the field $\bar{z}$ relaxes in the opposite time direction ($z$ and $\bar{z}$ also interact). In Figure~\ref{fig:contourplot} we show contour plots of $z(x,\tau)$ and $\bar{z}(x,\tau)$ 
for the saddle-point solution relevant to  calculating the purity (bipartite entanglement) of one half of the chain, where the boundary condition enforces a sharp domain wall in $z(x)$ at the final time.

Once we have established this picture for free fermions, we include weak interactions, which induce a crossover to a distinct universality class for entanglement generation. In the replica approach, the free-fermion model has a continuous ${\rm SO}(2N)$ symmetry 
which leads ``diffusive'' growth of entanglement. However, if we add weak 4--Majorana terms to our model (changing the free-fermion system into an interacting fermion system), the continuous symmetry is explicitly broken down to a discrete symmetry.
This may be taken into account in the saddle-point treatment.
The initially sharp domain wall enforced by the boundary conditions relaxes to some extent, but reaches a steady state at a finite length scale $l_\mathrm{int}$, set by the strength of the interactions. At length scales $l\gg l_\mathrm{int}$ this is indistinguishable from a sharp domain wall similar to those found for strongly interacting systems, indicating a crossover from free behaviour to interacting behaviour in entanglement generation.

The physical consequence of this is a crossover from ``diffusive'' to ``ballistic'' spreading of information. 
A bipartite entanglement entropy equal to  $S$ implies (by counting degrees of freedom in an interval of a given size) that correlations  exist over a length scale $l$ that is also of order $S$.
The diffusively relaxing domain wall leads to ${S \sim \sqrt{t}}$, 
so the length scale of correlations grows diffusively in the free case, 
whereas a sharp domain wall leads to 
${S \sim t}$, 
meaning the length scale grows ballistically, as in generic strongly interacting systems.
This ``ballistic'' spreading of information should not be confused with ballistic motion of quasiparticles (which are of course not well-defined in the interacting system).

In many non-random free systems it has been possible to understand the scaling of entanglement via an intuitive picture in terms of the trajectories of quasiparticles \cite{calabrese2005evolution,
fagotti2008evolution,
calabrese2009entanglement,
alba2018entanglement}.
The picture that we obtain here for the noisy system is different, but we will make a heuristic connection with the quasiparticle picture in Section \ref{sec:outlook}.

\tableofcontents

\section{Free model}

\subsection{Majorana chain} \label{sectionmajoranachain}

We focus on a 1D chain of Majorana modes interacting randomly with their nearest neighbours (as in \cite{bao2021symmetry}, but without measurements). We wish to study how different regions of the chain become entangled over time, starting from an initial state with short-range entanglement. 
Our main focus  will  be on the dynamics of the purity $\Tr[\rho_A^2]$
(see Sec.~\ref{sec:entanglemententropy})
for a region $A$ which we will usually take to be the right half of the chain.

There is a Majorana operator $\gamma_i$ for each site $i$, where $i$ ranges from $1$ to $L$ ($L$ even). 
The Majorana operators are Hermitian, ${\gamma_i^\dagger = \gamma_i}$, and obey the anti-commutation relations $\{\gamma_i,\gamma_j\}=2\delta_{ij}$.

The Hilbert space $\mathcal{H}$ has dimension $2^{L/2}$ and is spanned by the simultaneous eigenstates of the operators $i \gamma_{2j-1} \gamma_{2j}$, which have eigenvalues $\pm 1$. If we define $L/2$ pairs complex fermion operators $c=(\gamma_{2j-1}+i\gamma_{2j})/2$ and $c^{\dagger}=(\gamma_{2j-1}-i\gamma_{2j})/2$, we find $i \gamma_{2j-1} \gamma_{2j} = 2c^{\dagger}c -1$, so these are simply occupation number states. Alternatively, we can map the Majorana chain to a spin chain with $L/2$ sites and identify $i\gamma_{2j-1}\gamma_{2j} = Z_j$ and $i\gamma_{2j}\gamma_{2j+1} = X_j X_{j+1}$ (where $X$ and $Z$ are Pauli matrices).

We take our time-dependent Hamiltonian to be
\begin{equation} \label{majoranahamiltonian}
    H_\gamma(t) = -i \sum_i \eta_i(t) \gamma_i \gamma_{i+1}
\end{equation}
where the $\eta_i$ on each site are independent Gaussian noise terms with zero mean and with correlation function $\langle \eta_i(t)\eta_j(t')\rangle = \Delta^2 \delta_{ij} \delta(t-t')$.

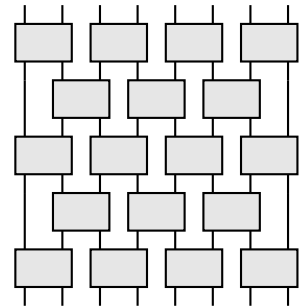
\begin{figure}
    \centering
    \begin{tikzpicture}
    \definecolor{grey}{gray}{0.9}
    \definecolor{blackfade}{gray}{0.5}
    \definecolor{greyfade}{gray}{0.95}
    \newcommand{\ufeet}[4]
    {
        \fill[#3] (#1-0.125,#2*0.75) rectangle (#1+0.625,#2*0.75+0.5);
        \draw[#4,thick] (#1-0.125,#2*0.75) rectangle (#1+0.625,#2*0.75+0.5);
        \draw[#4,thick] (#1,#2*0.75-0.25) -- (#1,#2*0.75);
        \draw[#4,thick] (#1+0.5,#2*0.75-0.25) -- (#1+0.5,#2*0.75);
    }
    \newcommand{\introw}[4]
    {
        \ufeet{#1}{#2}{#3}{#4}
        \ufeet{1+#1}{#2}{#3}{#4}
        \ufeet{2+#1}{#2}{#3}{#4}
        \ufeet{3+#1}{#2}{#3}{#4}
    }
    \newcommand{\halfrow}[4]
    {
        \ufeet{0.5+#1}{#2}{#3}{#4}
        \ufeet{1.5+#1}{#2}{#3}{#4}
        \ufeet{2.5+#1}{#2}{#3}{#4}
        \draw[thick,#4] (#1,#2*0.75-0.25) -- (#1,#2*0.75+0.5);
        \draw[thick,#4] (#1+3.5,#2*0.75-0.25) -- (#1+3.5,#2*0.75+0.5);
    }
    \newcommand{\toplegs}[3]
    {
        \draw[thick,#3] (#1,#2-0.25) -- (#1,#2);
        \draw[thick,#3] (#1+0.5,#2-0.25) -- (#1+0.5,#2);
        \draw[thick,#3] (#1+1,#2-0.25) -- (#1+1,#2);
        \draw[thick,#3] (#1+1.5,#2-0.25) -- (#1+1.5,#2);
        \draw[thick,#3] (#1+2,#2-0.25) -- (#1+2,#2);
        \draw[thick,#3] (#1+2.5,#2-0.25) -- (#1+2.5,#2);
        \draw[thick,#3] (#1+3,#2-0.25) -- (#1+3,#2);
        \draw[thick,#3] (#1+3.5,#2-0.25) -- (#1+3.5,#2);
    }
    \newcommand{\drawcircuit}[4]
    {
        \introw{#1}{#2}{#3}{#4}
        \halfrow{#1}{0.75+#2}{#3}{#4}
        \introw{#1}{1.5+#2}{#3}{#4}
        \halfrow{#1}{2.25+#2}{#3}{#4}
        \introw{#1}{3+#2}{#3}{#4}
        \toplegs{#1}{3.75+0.75*#2}{#4}
    }
    \drawcircuit{0}{0}{grey}{black}
    \end{tikzpicture}
    \caption{The unitary circuit corresponding to the Trotterized time evolution. Each grey rectangle is an independent two-site unitary of the form $u(\eta) = \exp(-\eta\gamma_i \gamma_{i+1})$.}
    \label{fig:brickwall}
\end{figure}

The time-evolution operator for a given realization of the noise is $U_\eta = \mathcal{T} e^{-i \int \dd t H_\gamma(t)}$, where $\mathcal{T}$ denotes time-ordering. For convenience we will Trotterize the time evolution to get a ``brick wall'' unitary circuit as shown in Figure \ref{fig:brickwall}, where the gates are independent random two-site unitaries of the form
\begin{equation}
    u(\eta) = \exp(-\eta\gamma_i \gamma_{i+1})
\end{equation}
and $\eta$ is a Gaussian random variable chosen independently for each gate with zero mean and variance
$\Delta^2 \delta t$. The number of double layers of the brickwall circuit is~$t/\delta t$.

\subsection{Replicas}
\label{sec:replicas}

For any given realization of noise terms $\eta$, the system undergoes unitary evolution $U_\eta$. 
We will be interested in the averages of physical quantities over realizations.

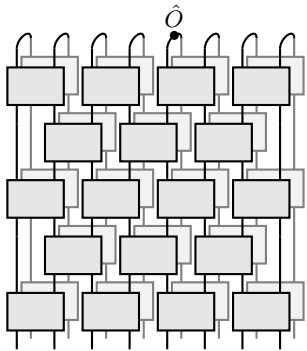
\begin{figure}
    \centering
    \begin{tikzpicture}
    \definecolor{grey}{gray}{0.9}
    \definecolor{blackfade}{gray}{0.5}
    \definecolor{greyfade}{gray}{0.95}
    \newcommand{\ufeet}[4]
    {
        \fill[#3] (#1-0.125,#2*0.75) rectangle (#1+0.625,#2*0.75+0.5);
        \draw[#4,thick] (#1-0.125,#2*0.75) rectangle (#1+0.625,#2*0.75+0.5);
        \draw[#4,thick] (#1,#2*0.75-0.25) -- (#1,#2*0.75);
        \draw[#4,thick] (#1+0.5,#2*0.75-0.25) -- (#1+0.5,#2*0.75);
    }
    \newcommand{\introw}[4]
    {
        \ufeet{#1}{#2}{#3}{#4}
        \ufeet{1+#1}{#2}{#3}{#4}
        \ufeet{2+#1}{#2}{#3}{#4}
        \ufeet{3+#1}{#2}{#3}{#4}
    }
    \newcommand{\halfrow}[4]
    {
        \ufeet{0.5+#1}{#2}{#3}{#4}
        \ufeet{1.5+#1}{#2}{#3}{#4}
        \ufeet{2.5+#1}{#2}{#3}{#4}
        \draw[thick,#4] (#1,#2*0.75-0.25) -- (#1,#2*0.75+0.5);
        \draw[thick,#4] (#1+3.5,#2*0.75-0.25) -- (#1+3.5,#2*0.75+0.5);
    }
    \newcommand{\toplegs}[3]
    {
        \draw[thick,#3] (#1,#2-0.25) -- (#1,#2);
        \draw[thick,#3] (#1+0.5,#2-0.25) -- (#1+0.5,#2);
        \draw[thick,#3] (#1+1,#2-0.25) -- (#1+1,#2);
        \draw[thick,#3] (#1+1.5,#2-0.25) -- (#1+1.5,#2);
        \draw[thick,#3] (#1+2,#2-0.25) -- (#1+2,#2);
        \draw[thick,#3] (#1+2.5,#2-0.25) -- (#1+2.5,#2);
        \draw[thick,#3] (#1+3,#2-0.25) -- (#1+3,#2);
        \draw[thick,#3] (#1+3.5,#2-0.25) -- (#1+3.5,#2);
    }
    \newcommand{\drawcircuit}[4]
    {
        \introw{#1}{#2}{#3}{#4}
        \halfrow{#1}{0.75+#2}{#3}{#4}
        \introw{#1}{1.5+#2}{#3}{#4}
        \halfrow{#1}{2.25+#2}{#3}{#4}
        \introw{#1}{3+#2}{#3}{#4}
        \toplegs{#1}{3.75+0.75*#2}{#4}
    }
    \newcommand{\topconnect}[3]
    {
        \draw[thick,#3] (#1,#2) .. controls (#1,#2+0.1875*0.75) and (#1+0.1875,#2+0.375*0.75) .. (#1+0.1875,#2+0.1875*0.75);
    }
    \newcommand{\topconnectdot}[4]
    {
        \draw[thick,#3] (#1,#2) .. controls (#1,#2+0.1875*0.75) and (#1+0.1875,#2+0.375*0.75) .. (#1+0.1875,#2+0.1875*0.75);
        \fill (#1+0.5*0.1875,#2+1.25*0.1875*0.75) circle (0.06cm) node[anchor=south]{#4};
    }
    \drawcircuit{0.1875}{0.1875}{greyfade}{blackfade}
    \drawcircuit{0}{0}{grey}{black}
    \topconnect{0}{3.75}{black}
    \topconnect{0.5}{3.75}{black}
    \topconnect{1}{3.75}{black}
    \topconnect{1.5}{3.75}{black}
    \topconnectdot{2}{3.75}{black}{$\hat{O}$}
    \topconnect{2.5}{3.75}{black}
    \topconnect{3}{3.75}{black}
    \topconnect{3.5}{3.75}{black}
    \end{tikzpicture}
    \caption{The folded unitary circuit for $\langle\psi\vert\hat{O}\vert\psi\rangle$. The front layer is the original unitary circuit $U_\eta$ and the back layer is its complex conjugate $U_\eta^*$. At the top, we show how a single-site observable $\hat{O}$ acts on the replicated system. At every other site, we have the single-site identity operator $\mathbb{1}$ denoted by an undecorated line.}
    \label{fig:foldedrep}
\end{figure}

The simplest quantity we can study is the expectation of some observable $\hat{O}$. If we prepare our system in the initial state $\vert\psi\rangle$ and this evolves over time into some new state $U_\eta\vert\psi\rangle$, 
then the averaged expectation value is $\overline{\langle\hat{O}\rangle} = \overline{\langle\psi\vert U_\eta^\dagger \hat{O} U_\eta \vert\psi\rangle}$, where the overline denotes the average over noise. 

Since $U_\eta$ and $U_\eta^\dagger$ are not independent, it is useful to 
``fold'' this expression as shown in Figure \ref{fig:foldedrep} to get a unitary circuit with two layers, one being the original circuit $U_\eta$ and the other being its complex conjugate $U_\eta^*$ (with respect to some orthonormal basis $\vert \psi_n \rangle)$. 
This double-layer circuit acts on two replicas of the original system, 
or in other words on a ket
${\vert\psi\rangle \otimes \vert\psi^*\rangle}$
in the replicated Hilbert space ${\mathcal{H} \otimes \mathcal{H}}$.
The expectation value $\langle\hat{O}\rangle$ is then a linear function of this ket,
so that the operator $\hat{O}$ is effectively a bra $\langle \hat{O} \vert$, and $\langle \psi \vert \hat{O} \vert \psi \rangle$ becomes ${\langle \hat{O} \vert (\vert\psi\rangle \otimes \vert\psi^*\rangle)}$. 
For example, the identity operator $\mathbb{1}$ becomes ${\langle \mathbb{1} \vert = \sum_n \langle \psi_n \vert \otimes \langle \psi_n \vert}$, where $\vert \psi_n \rangle$ are the basis vectors.
(Note that we will use Dirac notation $\ket{\bullet}$ for kets in both  physical and the replicated Hilbert spaces.)

The double-layer circuit is made up of gates of the form $u\otimes u^*$.
Taking the average over disorder is now straightforward because these gates  are independent from one another
for  different points in spacetime. To find the expectation value of some operator $\hat{O}$ averaged over noise, we simply start with the replicated initial state $\vert\psi\rangle \otimes \vert\psi^*\rangle$, evolve deterministically using the averaged replicated gates $\overline{u \otimes u^*}$, and then find the overlap with~${\langle \hat{O} \vert}$.

We can use a similar trick to calculate the average over noise of the $N$th moment of the expectation value,  $\overline{\langle\hat{O}\rangle^N}$, by considering a circuit with $2N$ layers acting on $2N$ replicas of the system. We can label these replicas with the index $a=1,\dots,2N$, where the original unitary circuit $U_\eta$ is applied for odd $a$ and its complex-conjugate $U_\eta^*$ for even $a$. In this case, the identity operator acting on every pair of replicas is
\begin{equation}
    \langle \mathbb{1} \vert = \Big( \sum_n \langle \psi_n \vert \otimes \langle \psi_n \vert \Big)^{\otimes N}.
\end{equation}

\subsection{Entanglement entropy}\label{sec:entanglemententropy}

Our main focus will be the way in which
quantum correlations are generated between different parts of the system by the random unitary dynamics.
We quantify this by the entanglement entropy of a subsystem $A$ ---  for example, the rightmost half of the chain ${i>L/2}$.  
If we prepare the chain in a state with short-range entanglement, this entropy will initially be small, but it will grow over time as the couplings entangle  different parts of the~chain.
Here we recall the expression for the R\'enyi entanglement entropies in the folded representation 
(readers familiar with this may wish to skip ahead).

For any integer ${N>1}$, the $N$th R\'enyi entanglement entropy  for a subsystem $A$ is defined as
\begin{equation} \label{renyientropy}
    S_N=-\frac{1}{N-1}\ln\left({\mathrm{Tr}[\rho^N_A]}\right)
\end{equation}
where $\rho_A$ is the reduced density matrix of the subsystem (the system as a whole is in a pure state). 
The limit $N\to 1$ gives the 
 von Neumann entanglement entropy.

The quantity $\mathrm{Tr}[\rho^N_A]$ does not correspond to the expectation value of any conventional observable because it is not linear in the reduced density matrix $\rho_A$. 
However, it does correspond to the expectation value of an operator acting on $N$ replicas of the system \cite{ekert2002direct,islam2015measuring}. 
This operator, which we denote by $C_A$,
performs a cyclic permutation of the replicas  within the subsystem $A$, while leaving everything outside the subsystem unchanged. 
(For simplicity, we will always assume that $A$ and $B$ each contain an even number of Majorana modes so that there is a well-defined Hilbert space in each subsystem.)

In the ``folded'' representation described above,  we now consider a $2N$-layer unitary circuit that acts on $2N$ replicas of the system. 
The expectation value of $C_A$ becomes a bra $\langle C_A \vert$,
with ${\mathrm{Tr}[\rho^N_A]=\langle C_A \vert ( \ket{\psi}\otimes\ket{\psi^*}\otimes\cdots \otimes \ket{\psi^*})}$.
The bra $\langle C_A \vert$ is a product state of the single-site identity bra $\langle\mathbb{1}\vert$ on every site in $B$, and the single-site bra $\langle C \vert$ 
which cycles replicas on every site in $A$:
\begin{equation}
    \langle C_A \vert = \bigotimes_i
    \begin{cases}
       \, \langle\mathbb{1}\vert \thickspace \,\,\, \mathrm{for}\thickspace i \notin A\\
      \,  \langle C \vert \thickspace \mathrm{for}\thickspace \,\, i \in A,
    \end{cases}
\end{equation}
where 
(taking $\langle \psi_n \vert$ to be an orthonomal single-site  basis)
\begin{equation}
\begin{split}
    \langle C \vert = \sum_{\{n_i\}}
    \langle \psi_{n_N} \vert \otimes \langle \psi_{n_1} \vert \otimes
    \langle \psi_{n_1} \vert \otimes \langle \psi_{n_2} \vert \otimes \\
    \cdots
    \otimes \langle \psi_{n_{N-1}} \vert \otimes \langle \psi_{n_N} \vert
\end{split}
\end{equation}

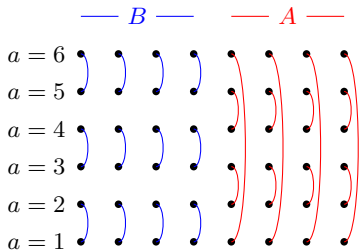
\begin{figure}
{
\centering
\begin{tikzpicture}
    \newcommand{\labgamma}[3]{\fill (#1,#2) circle (0.05cm) node[anchor=east] {#3};}
    \newcommand{\ungamma}[2]{\fill (#1,#2) circle (0.05cm);}
    \newcommand{\gammacol}[1]
    {
        \ungamma{#1}{0}
        \ungamma{#1}{0.5}
        \ungamma{#1}{1}
        \ungamma{#1}{1.5}
        \ungamma{#1}{2}
        \ungamma{#1}{2.5}
    }
    \newcommand{\icid}[1]
    {
        \draw (#1,0) .. controls (#1+0.125,0) and (#1+0.125,0.5) .. (#1,0.5);
        \draw (#1,1) .. controls (#1+0.125,1) and (#1+0.125,1.5) .. (#1,1.5);
        \draw (#1,2) .. controls (#1+0.125,2) and (#1+0.125,2.5) .. (#1,2.5);
    }
    \newcommand{\ictau}[1]
    {
        \draw (#1,0.5) .. controls (#1+0.125,0.5) and (#1+0.125,1) .. (#1,1);
        \draw (#1,1.5) .. controls (#1+0.125,1.5) and (#1+0.125,2) .. (#1,2);
        \draw (#1,0) .. controls (#1+0.25,0) and (#1+0.25,2.5) .. (#1,2.5);
    }
    \labgamma{0}{0}{$a=1\thickspace$}
    \labgamma{0}{0.5}{$a=2\thickspace$}
    \labgamma{0}{1}{$a=3\thickspace$}
    \labgamma{0}{1.5}{$a=4\thickspace$}
    \labgamma{0}{2}{$a=5\thickspace$}
    \labgamma{0}{2.5}{$a=6\thickspace$}
    \gammacol{0.5}
    \gammacol{1}
    \gammacol{1.5}
    \gammacol{2}
    \gammacol{2.5}
    \gammacol{3}
    \gammacol{3.5}
    \color{blue}
    \node at (0.75, 3) {$B$};
    \draw (0,3) -- (0.5,3);
    \draw (1,3) -- (1.5,3);
    \icid{0}
    \icid{0.5}
    \icid{1}
    \icid{1.5}
    \color{red}
    \node at (2.75, 3) {$A$};
    \draw (2,3) -- (2.5,3);
    \draw (3,3) -- (3.5,3);
    \ictau{2}
    \ictau{2.5}
    \ictau{3}
    \ictau{3.5}
    \color{black}
\end{tikzpicture}
\caption{An example of the index contractions involved in calculating the $N$th R\'enyi entropy of entanglement, here with $N=3$. 
Each row of sites,
labelled by the replica index $a$,
represents one the $2N=6$ replicas  arising from $\rho^{\otimes N}$.
Outside  subsystem $A$ (in its complement $B$), each replica is contracted with its conjugate. This takes the partial trace of each of the $N$ copies of the density matrix to give $\rho_A^{\otimes N}$.
Inside $A$, every even replica is contracted with the ``next'' replica to give $\rho_A^N$ and the last replica is contracted with first to give $\mathrm{Tr} [\rho_A^N]$. In the ``folded'' representation, this entire index contraction becomes a bra $\langle C_A\vert$.}\label{figindexcontractions}
}
\end{figure}

We can alternatively think of the operator $C_A$ as performing an index contraction, as illustrated in Fig.~\ref{figindexcontractions}  for the case $N=3$.
Outside of the subsystem $A$, the state $\bra{C_A}$ acts like the identity state $\bra{\mathbb{1}}$, 
which simply contracts each replica ${a=2m-1}$ with its complex conjugate ${a=2m}$ for all ${m=1\dots N}$
(generalizing the contractions at the top of Fig.~\ref{fig:foldedrep}).
Inside subsystem $A$, the operators $C_A$ permutes the odd indices in a cycle before contracting the replicas in these pairs, so it contracts $a=2m$ with $a=2m+1$ for $m=1\dots N-1$ and contracts $a=2N$ with $a=1$:
see Fig.~\ref{figindexcontractions}.

Finally, it is useful to characterize these states in the Majorana language.
In Appendix~\ref{appindexcontractionmajoranas} we show that if the bra $\langle C_A\vert$ contracts two replicas $a$ and $b$ of a Majorana chain at site $i$, this simply means that $\langle C_A\vert$ is an eigenstate of ${-i\gamma^a_i\gamma^b_i}$ with eigenvalue +1 (when $a<b$). 
Given that there are $N$ such independent stabilizers at each site, and the local Hilbert space dimension is $2^N$, these stabilizers uniquely determine the state $\langle C_A\vert$ up to a phase (the phase is then fixed by the fact that ${\mathrm{Tr}\rho^N_A}$ must be real and positive).
 
\section{Heisenberg mapping}

We now return to the noisy Majorana Hamiltonian. We show that taking the average over noise in the folded representation with $2N$ replicas  gives an ${\rm SO}(2N)$ Heisenberg model evolving in imaginary time, similarly to \cite{bao2021symmetry}.

\subsection{Average over noise}

First we need the form of the averaged gates. Using lower indices for sites and upper indices for replicas, the replicated gates take the form
\begin{equation}
    \tilde{u}(\eta) = \exp\left(-\eta\sum_{m=1}^{N}\lf \gamma_i^{2m-1}\gamma_{i+1}^{2m-1}-\gamma_i^{2m}\gamma_{i+1}^{2m}
\ri \right)
\end{equation}
where the minus sign comes from taking the complex conjugate on even replicas, 
in a basis where  
${i\gamma_i^a\gamma_{i+1}^a}$ are all real. 
(The choice of basis is a matter of convention, but this choice was natural given the standard Ising mapping
${i\gamma_{2j-1}\gamma_{2j} = Z_j}$ and ${i\gamma_{2j}\gamma_{2j+1} = X_j X_{j+1}}$.)

However, this minus sign is inconvenient, so we will absorb it into the definition of the Majorana operators. 
We  redefine the Majorana operators on even sites and even replicas via ${\gamma_{2j}^{2m} \mapsto -\gamma_{2j}^{2m}}$. 
This us gives us
\begin{equation} \label{enlargedsymmetry}
    \tilde{u}(\eta) = \exp\left(-\eta\sum_{a=1}^{2N}\gamma_i^a\gamma_{i+1}^a\right)
\end{equation}
This change reveals an ${\rm SO}(2N)$ symmetry under replica rotations which we discuss in section \ref{sssymmetrygroup}. 
This symmetry does not depend on the fact that the lattice is bipartite:\footnote{If we added next-nearest-neighbour interactions $i\gamma_{i}\gamma_{i+2}$, these terms would not be affected by the above sign change, but $i\gamma_{i}\gamma_{i+2}$ is pure imaginary so the ${\rm SO}(2N)$ symmetry is apparent even before the sign change.} ${\rm SO}(2N)$ symmetry arises for any quadratic Majorana Hamiltonian, and therefore for any free-fermion Hamiltonian, since complex fermions may always be expressed in terms of Majoranas. If the Hamiltonian has additional symmetries, such as charge conservation, the replica symmetry may be still larger.
(If the evolution becomes non-unitary, say due to measurements, then the symmetry is reduced \cite{bao2021symmetry} unless an appropriate bipartite structure is preserved \cite{sigmamodelmeasurement}.)

Taking the average over noise gives
\begin{align}
    \overline{\tilde{u}} &= \exp\left(\frac{\Delta^2\delta t}{2}\left[\sum_{a=1}^{2N}\gamma_i^a\gamma_{i+1}^a\right]^2\right) \\ &= \exp\left(-\Delta^2\delta t\left[N-\sum_{a<b}^{2N}A_i^{ab}A_{i+1}^{ab}\right]\right)
\end{align}
where we have defined the single-site Hermitian operators
\begin{equation}
    A_i^{ab}\equiv-\frac{i}{2}[\gamma_i^a,\gamma_i^b]
\end{equation}
These operators can be viewed as generators of the ${\rm so}(2N)$  Lie algebra (Sec.~\ref{sssymmetrygroup}). They commute on different lattice sites,
$[A_i^{ab},A_j^{cd}]=0$ if $i\neq j$, and square to the identity, 
$(A_i^{ab})^2=\mathbb{1}$.

In order to recover our original continuous-time model (\ref{majoranahamiltonian}), we take the limit ${\delta t \to 0}$. In this limit, the unitary gates acting during a given time-step effectively commute, so the (replicated and averaged) system evolves deterministically in imaginary time with the effective Hamiltonian
\begin{equation} \label{generalhamiltonian}
    H = \Delta^2\sum_i\left[N-\sum_{a<b}^{2N}A_i^{ab}A_{i+1}^{ab}\right].
\end{equation}
This is a ferromagnetic Heisenberg model, as we discuss in Sec.~\ref{sssymmetrygroup}. 
Before that, we briefly describe an alternative formulation of the averaged dynamics.

\subsection{Alternative stochastic description}
\label{sec:stochastic}

The dynamics generated by the Hamiltonian \eqref{majoranahamiltonian} and its discretized version can be viewed as a stochastic dynamics on the quantum many-body Hilbert space, as in the model systems discussed in \cite{Bauer2017Stochastic}. Let us elaborate briefly on this connection.

The stochastic unitary evolution $U_{t;t+\dd t}=U_{t+\dd t}U_t^{-1} $  between time $t$ and $t+\dd t$ on the Majorana Hilbert space generated by the Hamiltonian \eqref{majoranahamiltonian} reads $U_{t;t+\dd t} = e^{-i\dd H_t}$ with Hamiltonian increment $\dd H_t$,  
\begin{equation} \label{eq:hamilton-dt}
\dd H_t = \sum_j E_j\, \dd B^j_t,
\end{equation}
with $E_j=-i\gamma_j\gamma_{j+1}$ and where $\dd B^j_t$ are Brownian increments normalized to $\dd B^j_t \dd B^k_t=\delta_{j,k}\Delta^2\, \dd t$ (formally $\dd B^j_t=\int_t^{t+\dd t}\eta_j(s)\dd s$).

The quantum state or density matrix $\rho_t$ evolves during the time interval ${[t,t+\dd t]}$ according to ${\rho_{t+\dd t} = e^{-i\dd H_t} \rho_t e^{+i\dd H_t}}$, so that its increment ${\dd\rho_t :=\rho_{t+\dd t}-\rho_t}$ is, up to order $O(\dd t^{3/2})$,
\begin{equation} \label{eq:EOM-rho}
\dd\rho_t = -i[\dd H_t,\rho_t] -\frac{1}{2} [\dd H_t,[\dd H_t,\rho_t]] .
\end{equation}
This way of writing the equation of motion implicitly assumes that the Brownian increments $\dd B^j_t$ are sampled from time $t$ to $t+\dd t$, independently of the evolution of the density matrix $\rho_t$ up to time $t$. This is the It\^o convention for stochastic calculus. This is also the convention taken in the discretized random unitary circuit of Fig.~\ref{fig:brickwall} in which the Gaussian random variables $\eta_j$, with variances $\Delta^2\delta t$ and independently updated at each step, represent the Brownian increments $\dd B_t^j$.  

The first term in \eqref{eq:EOM-rho} is a purely noisy contribution, since the Hamiltonian increment \eqref{eq:hamilton-dt} contains no deterministic part, whereas by the It\^o rule ${\dd B_t^j\dd B_t^k=\delta_{jk}\Delta^2 \dd t}$ mentioned above, the second term is a purely deterministic drift. Since $\dd H_t$ and $\rho_t$ are independent and $\dd H_t$ has zero mean, the first term in \eqref{eq:EOM-rho} vanishes in average, and the time evolution of the mean density state $\overline{\rho_t}$ possesses a standard Lindblad form and reads
\begin{equation}
    \dd\overline{\rho_t} = -\frac{1}{2} [\dd H_t,[\dd H_t,\overline{\rho_t} ]] =: \mathcal{L}(\overline{\rho_t})\ \dd t, 
\end{equation}
where the last equality defines the (one-replica) Lindblad operator $\mathcal{L}$.

A similar structure also emerges when looking at higher moments of the density matrix, and hence at higher replicas. The $N$-times replicated state $\rho_t^{\otimes N}$ evolves randomly but unitarily with
\begin{equation}
\rho_{t+\dd t}^{\otimes N} = e^{-i\dd H_t^{(N)}}\, \rho_t^{\otimes N}\, e^{+i\dd H_t^{(N)}},
\end{equation}
with $N$th replica Hamiltonian increments ${\dd H_t^{(N)}=\sum_j E_j^{(N)}\, \dd B^j_t}$,
where $E_j^{(N)}=\sum_\alpha E_j^\alpha$, and ${\alpha}$ (which runs over ${N}$ values,
in contrast to the indices $a,b$ in  in Eq.~(\ref{generalhamiltonian}) which run over $2N$ values)
indexes the replicas of the density matrix.\footnote{Alternatively, 
$E_j^{(N)}=E_j\otimes\mathbb{1}\cdots\otimes\mathbb{1} + \cdots +\mathbb{1}\otimes \cdots\mathbb{1}\otimes E_j $, with the tensor structure reflecting the replicas.} 
Each of the replicas is coupled identically to the noise through the operators $E_j$, 
and the replicas are non-interacting before any average.

As a consequence, the replicated density matrix $\rho_t^{\otimes N}$ evolves according to the SDE \eqref{eq:EOM-rho} but with $\dd H_t$ replaced with $\dd H_t^{(N)}$. Taking then the average, we learn that 
\begin{equation}
    \dd \overline{\rho_t^{\otimes N}} = -\frac{1}{2} [\dd H_t^{(N)},[\dd H_t^{(N)},\overline{\rho_t} ]] =: \mathcal{L}^{(N)}(\overline{\rho_t^{\otimes N}})\ \dd t.
\end{equation}
The last equality defines the $N$-replica Lindblad operator:
\be
\mathcal{L}^{(N)}(\bullet )
= - \f{\Delta^2}{2} \sum_j \sum_{\alpha,\alpha'} 
\left[ E^\alpha_j, \left[E^{\alpha'}_j, \, \bullet \, \right] \right].
\ee 
By construction, $\mathcal{L}^{(N)}$  coincides with the Hamiltonian \eqref{generalhamiltonian} which is thus a Lindblad operator (but acting on density matrices on the physical Hilbert space replicated $N$ times).

Actually, the equation of motion \eqref{eq:EOM-rho} is simply a classical stochastic differential equation (SDE) but on many-body quantum density matrices. Like any SDE, it is associated with a Fokker-Planck operator. Due to the linearity of \eqref{eq:EOM-rho}, this Fokker-Planck operator preserves polynomials
in $\rho$ of any fixed  degree, and its restriction to monomials of order $N$ is the  Lindblad operator $\mathcal{L}^{(N)}$, or equivalently $H$ as in \eqref{generalhamiltonian}.

By an argument similar to that used in the case of noisy complex fermionic chains \cite{Bauer2019Equilibrium}, this description leads to a simple understanding of the invariant probability distribution, reached at large time (for finite $L$). Indeed, the operators $E_j$ generate the Lie algebra ${\rm so}(L)$ by iterative commutation relations. These multiple commutators form a representation of ${\rm so}(L)$ on the  Majorana Fock space which decomposes into two irreducible ${\rm so}(L)$ representations depending on the chirality ($L$ is even). As a consequence, iterations of multiple products $e^{-i\dd H_{t_k}}$, for series of times $t_1,\ t_2,\cdots, t_p$, that is
\[ 
e^{-i\dd H_{t_1}}e^{-i\dd H_{t_2}}\cdots e^{-i\dd H_{t_p}},
\]
span the group ${\rm SO}(L)$.\footnote{More precisely, they form a dense set of operators on the image of ${\rm SO}(L)$ by the representation map on the Majorana Fock space.} Since the invariant measure is, by definition, stable under the dynamics generated by $\dd H_t$, it is ${\rm SO}(L)$ invariant. Since the ${\rm SO}(L)$ action is faithful on its spin representations, this invariant measure is thus identical to the measure induced by the ${\rm SO}(L)$ Haar measure on each of the two irreducible ${\rm SO}(L)$ components of the Majorana Fock space.

\subsection{Symmetry group} \label{sssymmetrygroup}

Let us return to the effective Hamiltonian (\ref{generalhamiltonian}).
The operators $A^{ab}$ are a generalization of Pauli matrices in that we can define the operators $J^{ab}\equiv\frac{1}{2}A^{ab}$ which have eigenvalues $\pm\frac{1}{2}$ and have angular momentum commutation relations
\begin{equation} \label{eq:amcommutationrelations}
    [J^{ab},J^{cd}] = i(\delta_{ac}J^{bd}+\delta_{bd}J^{ac}-\delta_{ad}J^{bc}-\delta_{bc}J^{ad}).
\end{equation}
The $J^{ab}$ therefore function as angular momentum operators in $2N$-dimensions, i.e. they form a representation of an ${\rm so}(2N)$ Lie algebra. The operator $J^{ab}$ generates rotations in the $ab$-plane.
    
The Hamiltonian in Eq.~(\ref{generalhamiltonian}) is  a ferromagnetic  Heisenberg model, with ${\rm SO}(2N)$ global symmetry \cite{bao2021symmetry}. The constant term ensures that the ground state energy of the model is zero. This is necessary:
as we will see below, expectation values in the Majorana chain map to transition amplitudes in the ferromagnet, and the existence of a state with negative energy would imply that expectation values could grow indefinitely over time, even in a finite system.
We also know that the state $\langle \mathbb{1} \vert$ corresponding to the identity operator satisfies $\langle \mathbb{1} \vert e^{-Ht}=\langle \mathbb{1} \vert$, meaning it is an energy eigenstate with zero energy and therefore a ground state. The ${\rm SO}(2N)$ symmetry mentioned above means that for $N>1$ the number of ground states increases with $L$, and these ground states approach a continuous manifold of states in the thermodynamic limit. 
The existence of this continuous manifold of ground states 
is in stark contrast to generic interacting systems, where the discrete replica symmetry guarantees only a discrete family of ground states, as we discuss in Sec.~\ref{sec:symmetriesgeneral}.

For each site on the original Majorana chain, the replicated system has $2N$ Majorana operators  (one for each replica), so the local Hilbert space dimension is $2^N$.  
Viewed as matrices on this space, the generators $A^{ab}$ therefore form a $2^N$-dimensional representation of ${\rm so}(2N)$. This is the bispinor representation of ${\rm so}(2N)$ (the Euclidean equivalent of a Dirac spinor in $2N$ dimensions).
This representation is reducible: it is the direct sum of two chiral spinor representations (the equivalent of Weyl spinors).
These $2^{N-1}$-dimensional representations are irreducible. (As representations of ${\mathrm{SO}(2N)}$, they are projective.) We can define the chirality operator on a site
\begin{equation}
    \chi = (-i)^N\prod_{a=1}^{2N}\gamma^a
\end{equation}
which has eigenvalues $\pm 1$ and commutes with the generators $J^{ab}$ (here we multiply the Majorana operators from left to right $\gamma^1 \gamma^2 \gamma^3 \dots$). States with $\chi=+1$ transform under one of the spinor representations, while states with $\chi=-1$ transform under the other representation.

\subsection{Special cases}

We can understand the implications of the mapping to the ferromagnetic chain by focussing on some special cases:

\subsubsection{$N=1$: Expectation values}\label{sec:expectationvalues}

The simplest case is $N=1$. This tells us about the expectation values of observables averaged over noise. In the folded representation, the unitary circuit has two layers labelled $a=1,2$, and the ket for the replicated system represents the physical  density matrix.

The local Hilbert space dimension of the chain is ${2^N=2}$ and there is only one generator $A_i^{12}\equiv A_i$ at each site, with eigenvalues $\pm 1$. We can think of there being a spin-$\frac{1}{2}$ at each site $i$, with $A_i$ acting like the Pauli matrix $\sigma_i^z$. The Hamiltonian
\begin{equation} \label{isinghamiltonian}
    H = \Delta^2\sum_i\left[1 - A_i A_{i+1}\right]
\end{equation}
then simply describes an Ising model with no applied fields.

Say we prepare our Majorana chain in the initial state $\vert\psi\rangle$ and want to calculate the average over noise of some operator $\overline{\langle\hat{O}(t)\rangle}$.
By the discussion in Sec.~\ref{sec:replicas},
this is given by
${\langle \hat{O} \vert e^{-Ht}\vert\tilde{\psi}\rangle}$,
where 
$\vert\tilde{\psi}\rangle=\vert\psi\rangle\otimes\vert\psi^*\rangle$
is the replicated initial state
(and $\vert\psi^*\rangle$  is  complex conjugated with respect to our fixed basis, in which ${i \gamma_i \gamma_{i+1}}$ is real).

Instead of thinking about this in the ``Schrodinger picture'', where we evolve the replicated initial state $\vert\tilde{\psi}\rangle$ to get $e^{-Ht}\vert\tilde{\psi}\rangle$, 
it is more convenient to think about it in the ``Heisenberg picture'', in which we evolve the bra,  ${\langle\hat{O}\vert\rightarrow\langle\hat{O}\vert e^{-Ht}}$. 
This has the advantage that once we have calculated $\langle\hat{O}\vert e^{-Ht}$, we can easily calculate $\overline{\langle\hat{O}(t)\rangle}$ for any initial state.

For this to be useful, we need a way of expressing $\langle\hat{O}\vert$ in terms of Ising configurations (i.e. simultaneous eigenstates of $\{ A_i\}$). In Appendix \ref{appexpectationvaluesising}, we show that if the operator $\hat{O}$ is a product of distinct Majorana operators, then $\langle\hat{O}\vert$ corresponds to a single Ising configuration, with $A_i=-1$ if $\gamma_i$ appears in $\hat{O}$, and $A_i=+1$ otherwise.

From this it is clear that all such states $\langle\hat{O}\vert$ decay exponentially over time except the two ground states of the Ising model, which have zero energy and are therefore unaffected by time evolution. The ground state with all $A_i=+1$ corresponds to the identity operator $\mathbb{1}$, which simply measures the normalization of the state $\langle\psi\vert\psi\rangle$. The ground state with all $A_i=-1$ corresponds to the fermion parity operator $U_p = (-i)^{L/2}\prod_i \gamma_i$, which is also conserved by the dynamics.

All other operators decay exponentially over time. For example the expectation value $\langle i\gamma_i\gamma_{i+1} \rangle$ corresponds to the Ising state $\langle\uparrow\cdots\uparrow\downarrow\downarrow\uparrow\cdots\uparrow\mid$ which has energy $4\Delta^2$, so its noise average will decay over time from its initial value as $\overline{\langle i\gamma_i\gamma_{i+1} \rangle_t}=e^{-4\Delta^2 t}\langle i\gamma_i\gamma_{i+1} \rangle_0$. The expectation value of $\langle i\gamma_i\gamma_{i+2} \rangle$ corresponds to $\langle\uparrow\cdots\uparrow\downarrow\uparrow\downarrow\uparrow\cdots\uparrow\mid$ which has energy $8\Delta^2$ so $\overline{\langle i\gamma_i\gamma_{i+2} \rangle_t}=e^{-8\Delta^2 t}\langle i\gamma_i\gamma_{i+2} \rangle_0$.

\subsubsection{$N=2$: Purity}

A more interesting case, and the focus of the rest of this paper, is $N=2$. This will allow us to calculate the averaged purity of a subsystem
\begin{equation}
    \overline{\Tr[\rho_A^2]} = \overline{e^{-S_2}}
\end{equation}
where  $S_2$ is the second R\'enyi entropy of entanglement.
To simplify the notation, let us define the quantity
\be\label{eq:definecheckS}
\check{S}_2 = - \ln \overline{\Tr[\rho_A^2]}.
\ee
Here the average of the purity is taken before taking the log, which means $\check{S}_2\neq\overline{S_2}$ in general.
However, we fill find empirically that fluctuations in $S_2$ are sufficiently small that the two kinds of average coincide to leading order  
(see comment at the end of this section).

The unitary circuit now has $2N=4$ layers, and the local Hilbert space dimension of the replicated system is $2^N=4$.
The generators $A_i^{ab}$ form a representation of the Lie algebra ${\rm so}(4)$ at each site, which is equal to ${\rm su}(2)\oplus {\rm su}(2)$. The ``spins'' at each site are Dirac spinors, composed of two Weyl spinors with chirality $\chi=\pm 1$.  These 2-component Weyl spinors
are nothing but regular spin-$\frac{1}{2}$ spins.
In other words, the four-dimensional Hilbert space at a site  splits into 
two doublets, 
each of which is a spin-$\frac{1}{2}$  under  one of the two $\mathrm{SU}(2)$ groups, and invariant under the other $\mathrm{SU}(2)$ (see Appendix \ref{app:symmetrygroupreduction}).

To calculate the purity of a subsystem $A$ after time $t$, we can start with the state $\langle C_A\vert$ (corresponding to the operator which swaps replicas within the subsystem), evolve for imaginary time $t$ using the Hamiltonian (\ref{generalhamiltonian}), and finally overlap the resulting state with the some replicated initial state $\vert\tilde{\psi}\rangle$. Let us consider these three elements one at a time.

The state $\langle C_A\vert$ corresponds to two different local operations: outside the subsystem $A$, it acts like the identity operator, while inside $A$ it swaps replicas. In Appendix \ref{appindexcontractiondomainwall} we show that both of these local operations map to spins with $\chi=+1$. Given that such spins transform non-trivially only under ${\rm so}(3)$ generators, we can think of them as having directions in 3-dimensional space. We choose generators such that the identity operator maps to spin ``up'' $\langle\uparrow\mid$, while the swap operator maps to spin ``right'' $\langle\rightarrow\mid$.

The Hamiltonian in eq. (\ref{generalhamiltonian}) commutes with the chirality operator $\chi_i$ on every site, so each $\chi_i$ is individually conserved. This means that if the initial state of the replicated system has $\chi_i=+1$ on every site, then this will be true at all times, and we can restrict the Hilbert space to these states of positive chirality. These spins have a local Hilbert space dimension of 2 and transform non-trivially under only one set of ${\rm so}(3)$ generators. The Hamiltonian (\ref{generalhamiltonian}) becomes:
\begin{equation} \label{heisenberghamiltonian}
    H = 2\Delta^2\sum_i\left[1-\vec{\sigma}_i\cdot\vec{\sigma}_{i+1}\right]
\end{equation}
where $\vec{\sigma_i}=(\sigma_i^x,\sigma_i^y\sigma_i^z)$ are the Pauli matrices for the positive chirality spins. This is simply a standard ferromagnetic ${\rm su}(2)$ Heisenberg model.

Finally, we need the replicated initial state $\vert\tilde{\psi}\rangle$. For simplicity, we will choose a specific initial state with short-range entanglement. We choose the state which is a simultaneous eigenstate of $i\gamma_{2j-1}\gamma_{2j}$ with eigenvalue $+1$ for all $j$. In this state, the Majorana modes are maximally entangled in pairs. In Appendix \ref{appreplicatedinitialstate} we show that replicating this state, and projecting onto the subspace with $\chi_i=+1$, gives a product state of entangled pairs of the form 
\be
\vert \tilde{\psi}^+ \rangle 
\equiv \, 
(\mid\uparrow\uparrow\rangle\ +\mid\downarrow\downarrow\rangle)^{\otimes L/2}.
\ee
Note that this state is invariant under rotations in the $xz$-plane, i.e. rotations generated by $\sum_i\sigma_i^y$.

Putting this all together we get
\begin{equation} \label{purityequation}     \overline{\Tr[\rho_A^2]} = e^{-\check{S}_2} = \langle C_A\vert e^{-Ht} \vert\tilde{\psi}^+\rangle
\end{equation}
where $\langle C_A\vert$ is a product state of $\langle\uparrow\mid$ outside subsystem $A$ and $\langle\rightarrow\mid$ inside subsystem $A$. If we take the subsystem $A$ to be the rightmost half of the system, we can write this schematically as $\langle C_A \vert = \langle\uparrow\cdots\uparrow\rightarrow\cdots\rightarrow\mid$.

\subsubsection{Higher replicas}

As mentioned above, the quantity $\check{S}_2$ is not generally equal to the average second R\'enyi entropy $\overline{S_2}$, given that we are taking the average over disorder of $e^{-S_2}$ before taking the log. To learn more about the distribution of $S_2$ we could calculate higher moments of the purity, of the form 
\be\label{eq:highermoments}
\overline{e^{-k S_2}}=
\overline{\Tr[\rho_A^2]^k}.
\ee
This is the generating function for $S_2$.
For integer ${k\geq 1}$, it can be written using $2kN$ replicas of the Majorana chain, 
which can be mapped to the ${\rm SO}(2kN)$ Heisenberg model in Eq.~(\ref{generalhamiltonian}).

However, as explained later, numerical evidence suggests $\check{S}_2$ and $\overline{S_2}$ are in fact equal to leading order in $t$. This means that many of the results we derive for $\check{S}_2$ will also apply to $\overline{S_2}$.
The replica approach could also be used to compute R\'enyi entropies $S_n$ for other $n$, but in this paper we will stick to $S_2$.

\section{Entanglement purity: Saddle point approximation}

Eq.~(\ref{purityequation}) tells us that 
the growth in entanglement of a subsystem (or the decrease in its purity) 
is given by an imaginary-time transition amplitude in the Heisenberg chain.

In the state $\langle C_A\vert$, the Heisenberg chain is locally in a ground state both inside and outside of subsystem $A$, but with sharp domain walls at the boundaries of $A$. Imaginary-time evolution suppresses high energy states, so after sufficiently long times the evolved state $\langle C_A\vert e^{-Ht}$ should be locally close to a ground state everywhere (but no longer normalized). The ground states of a Heisenberg ferromagnet are product states of identical spins (or ``cat'' states formed from superpositions of such states). We therefore expect that, after sufficiently long times, the state is close to a product state, with the spin direction slowly varying with position. Over time, this product state should slowly relax to become smoother and smoother, and in a finite system should approach a uniform spin direction.

This intuitive picture suggests that, for long times ${\Delta^2 t \gg 1}$, quantum fluctuations are small for most of the trajectory. This motivates applying a saddle-point approximation to Eq.~(\ref{purityequation}).

In fact, it turns out that thinking in terms of $\langle C_A\vert e^{-Ht}$ as a product state evolving semiclassically is too simplistic. 
A saddle-point calculation is indeed valid in the regime ${\Delta^2 t\gg 1}$,
but it requires us to consider both ``forward'' and ``backward'' evolving fields as discussed below.

\subsection{Coherent state path integral}

In order to use a saddle-point approximation, we formulate the right-hand side of (\ref{purityequation}) as a coherent state path integral following Refs.~\cite{stone2000semiclassical,tailleur2008mapping}. We define single-spin coherent states in terms of the complex variable $z$
\begin{equation} \label{spincoherentstate}
    \vert z\rangle = (1+z^* z)^{-\frac{1}{2}}\left(\begin{matrix}z\\1\end{matrix}\right)
\end{equation}
which is just a stereographic projection from the Bloch sphere to the complex plane. We have chosen a basis such that the ``south pole'' of the Bloch sphere corresponds to $z=0$ and the $xz$-plane corresponds to real $z$. A spin with azimuthal angle $\theta$ and polar angle $\phi$ is described by $z=e^{-i\phi}\cot\frac{\theta}{2}$.

We will first recall \cite{stone2000semiclassical} how the coherent state path integral for a single spin is used to compute the  coherent state propagator: 
\begin{equation} \label{singlespinpropagator}
    K(z_I,z_F^*,t) = \langle z_F \vert e^{-Ht} \vert z_I \rangle.
\end{equation}
The single-spin identity operator may be resolved as
\begin{equation}
    \mathbb{1}=\int\mathrm{d}\mu\vert z \rangle \langle z \vert,
\end{equation}
where the measure
\begin{equation}\label{eq:measure}
    \mathrm{d}\mu=\frac{2}{\pi}\frac{\mathrm{d}^2 z}{(1+\bar{z}z)^2}
\end{equation}
is uniform over the Bloch sphere. 
Here $\mathrm{d}^2 z$ is $\dd a \dd b$ where $z=a+bi$ and $\bar{z}=a-bi$.
By inserting many such resolutions into the right-hand side of (\ref{singlespinpropagator}), we get a path integral form for the propagator
\begin{equation}
    K(z_I,z_F,t) = \int_{z(0)=z_I}^{\bar{z}(t)=z_F^*} D\mu \exp(-\mathcal{S})
\end{equation}
with an action $\mathcal{S}$  given by
\begin{equation}\label{eq:totalaction}
    \mathcal{S} = -\frac{1}{2}\int\mathrm{d}\tau\frac{\dot{\bar{z}}z-\bar{z}\dot{z}}{1+\bar{z}z}+\int\mathrm{d}\tau H(z,\bar{z}) + \mathcal{S}_\mathrm{bdry}
\end{equation}
This action contains a Berry phase term, a term resulting from the Hamiltonian $H(z,\bar{z}) \equiv \<z|H|z\>$, and a term $\mathcal{S}_\mathrm{bdry}$ from the initial and final-time boundaries given by
\begin{equation}
    \mathcal{S}_\mathrm{bdry} = -\frac{1}{2}\ln\frac{(1+z_F^* z(t))(1+\bar{z}(0)z_I)}{(1+z_F^* z_F)(1+z_I^* z_I)}.
\end{equation}
When this path integral 
is treated via saddle-point,
it is crucial to treat 
$z$  and
$\bar{z}$ 
as independent 
in searching for saddle-point trajectories, 
so that in general $\bar{z}\neq z^*$ \cite{stone2000semiclassical}  (or equivalently  $a$ and $b$ as defined below Eq.~(\ref{eq:measure}) become complex).  
Note that boundary conditions are fixed at opposite boundaries for the two fields:
$z$ is fixed only at $\tau=0$ by the initial state $\vert z_I \rangle$, so that $z(0)=z_I$, whereas $\bar{z}$ is fixed only at $\tau=t$ by the final state $\langle z_F \vert$, so that $\bar{z}(t)=z_F^*$.

The above formulation is for a single spin, but can be extended immediately to multiple spins. We simply sum over Berry phase terms and boundary terms in the action, and replace the single-spin Hamiltonian with the full many-body Hamiltonian from eq. (\ref{heisenberghamiltonian}). In the continuum limit (taking the lattice spacing to be $1$ and replacing $\sum_i$ with $\int \mathrm{d}x$) this gives us
\begin{equation} \label{continuumaction}
    \mathcal{S} = -\frac{1}{2}\int\mathrm{d}x\mathrm{d}\tau\frac{\dot{\bar{z}}z-\bar{z}\dot{z}}{1+\bar{z}z}+\int\mathrm{d}\tau H(z,\bar{z}) + \mathcal{S}_\mathrm{bdry}
\end{equation}
where the continuum Hamiltonian $H(z,\bar{z})$ is given by
\begin{equation}
    H(z,\bar{z}) = 4\Delta^2 \int\mathrm{d}x \frac{\bar{z}'z'}{(1+\bar{z}z)^2}
\end{equation}
(with $'$ denoting a spatial derivative, and suppressing arguments whenever possible) and the boundary term is now
\begin{equation}
\mathcal{S}_\mathrm{bdry} = -\frac{1}{2}\int\mathrm{d}x\ln\frac{(1+z_F^* z(t))(1+\bar{z}(0)z_I)}{(1+z_F^* z_F)(1+z_I^* z_I)}.
\end{equation}

\subsection{Equations of motion}

In general, this path integral cannot be evaluated exactly. However,  the intuitive argument above suggests that when $\Delta^2 t\gg 1$, we should be able to make use of a saddle-point approximation, and  indeed this  will be justified below (Sec.~\ref{sec:saddleptvalidity}). 
In the saddle-point approach we consider the single trajectory for which action is stationary and approximate
\begin{equation}
    K(z_I,z_F^*,t) \approx K_\mathrm{fl}(z_I,z_F^*,t) e^{-\mathcal{S}}
\end{equation}
where $\mathcal{S}$ is the action of the stationary trajectory and $K_\mathrm{fl}$ comes from integrating over quadratic fluctuations around this stationary trajectory.
Here the arguments $z_I$ and $z_F$ that specify the coherent states at the initial and final times are functions of position, $z_I(x)$ and~$z_F(x)$.

If we vary the action $\mathcal{S}$ while keeping the boundary conditions fixed and set this variation to zero, we find that the trajectory must obey the  equations of motion
\begin{align}
    \dot{z} &= + 4\Delta^2 \left( z'' - \frac{2\bar{z}(z')^2}{1+\bar{z}z} \right), \label{eom1}\\
    \dot{\bar{z}} &= - 4\Delta^2 \left( \bar{z}'' - \frac{2z(\bar{z}')^2}{1+\bar{z}z} \right), \label{eom2}
\end{align}
with the boundary conditions $z(x,0)=z_I(x)$ and  $\bar z(x,t)=z_F^*(x)$.

These equations look like diffusion equations with extra terms. Note that, due to the opposite signs in the two equations, $z$ diffuses in the positive time direction, while $\bar{z}$ diffuses in the negative time direction.

\subsection{Boundary conditions and the classical trajectory}

The equations of motion (\ref{eom1}) and (\ref{eom2}) enforce local constraints on the trajectory, but to find the unique classical trajectory we must consider the initial and final boundaries.

Given that ${\langle C_A \vert e^{-Ht} \vert \tilde{\psi}^+ \rangle = \langle \tilde{\psi}^+ \vert e^{-Ht} \vert C_A \rangle}$, we 
are free to consider 
our ``initial'' state at ${\tau=0}$ to be $\vert C_A \rangle$
and our ``final'' state at ${\tau=t}$ to be ${\vert \tilde{\psi}^+ \rangle}$.
That is, we will use a  coordinate $\tau$ 
that is \textit{reversed} with respect to physical time.
This is convenient because the state $\vert C_A \rangle = {\mid\uparrow\cdots\uparrow\rightarrow\cdots\rightarrow\rangle}$ will impose a sharp domain-wall boundary condition for $z(x,\tau)$ at $\tau=0$.
 
Since the symmetry between $\ket{\uparrow}$ and $\ket{\rightarrow}$ 
is not immediately apparent in the coherent states parameterization 
(\ref{spincoherentstate}),
it will  be convenient to make a symmetry rotation of our coherent states basis so that the domain wall boundary condition at $\tau=0$ becomes  $z=-\tan(\pi/8)$ on the left and $z=+\tan(\pi/8)$ on the right.

While the initial state $\vert C_A \rangle$ is already a product of coherent states, and translates directly into a boundary condition for $z(x,\tau)$ at ${\tau=0}$,
the final state $\vert \tilde{\psi}^+ \rangle$ is not (it is a product of entangled pairs of spins). 
Therefore we cannot immediately employ the above result for the propagator (\ref{singlespinpropagator}). 
However, we can evaluate the overlap $\langle \tilde{\psi}^+ \vert e^{-Ht} \vert C_A \rangle$ 
by writing it as
 $\int \dd \mu_w \langle \tilde{\psi}^+ \vert w \rangle \langle w \vert e^{-Ht} \vert C_A \rangle$, 
 where $\ket{w}$ denotes a product of coherent states with parameters $w_i$, and $\dd\mu_w$ is the product of the measures for these states.  Treating $w_i$, $\bar w_i$ as extra degrees of freedom in the path integral, we see that  they become the boundary values of the fields, $w_i=z_i(t)$ and $\bar w_i=\bar z_i(t)$.
We can absorb the extra factor from the overlap with ${\langle\tilde \psi^+|}$ into the action~via
\be\label{eq:boundaryconstraintPsi}
\langle \tilde{\psi}^+ \vert z(t) \rangle 
=
\exp \Big[ \Psi(z(t)) - \frac{1}{2}\sum_i\ln(1+\bar{z}_i(t) z_i(t))\Big],
\ee
where the function ${\Psi(w)=\Psi(w_1,\ldots w_L)}$ depends only on $w$ (not $\bar w$), 
and the second term comes from the normalization of the coherent state.

Varying the action including these terms, the bulk saddle point equations (\ref{eom1},~\ref{eom1}) are supplemented by the boundary equation 
\begin{equation} \label{generaltauequalstboundary}
    \frac{\partial \Psi(z(t))}{\partial z_i(t)} = \frac{\bar z_i(t)}{1+\bar z_i(t) z_i(t)},
\end{equation}
which comes from variation with respect to $\bar z_i(t)$.
This equation provides a relationship between the boundary values  $z(t)$ and $\bar z(t)$ for the two fields at $\tau=t$,
given the function $\Psi(w)$, which is in turn determined by the boundary state~${\vert \tilde{\psi} \rangle}$.

In the present case the boundary condition at $\tau=t$ (which corresponds to the \textit{physical} initial time)
becomes remarkably simple.
In Appendix \ref{appfinaltimeboundary} we show that our specific  initial state $\vert \tilde{\psi} \rangle$ has the effect of swapping the spins on neighbouring sites at the $\tau=t$ boundary. More precisely, the boundary constraint (\ref{generaltauequalstboundary}) is solved by setting
\ba
\bar z_{2j-1}(t) & =z_{2j}(t),
& 
\bar z_{2j}(t) & =z_{2j-1}(t). 
\end{align}
The swapping of the spatial indices here 
is significant at microsopic timescales.
But at late times ${\Delta^2 t \gg 1}$ the spins are slowly varying with position, 
and it is sufficient to approximate the equations above as\footnote{The boundary condition (\ref{eq:reflectingBC}) simply neglects the specific pattern of correlations found in the initial state $\vert\psi\rangle$, which we have arbitrarily chosen to be $i\gamma_{2j-1}\gamma_{2j}\vert\psi\rangle=+\vert\psi\rangle$. This does not affect the asymptotic entanglement growth at late times, which will be the same for any initial state with only short-range correlations.}
\be\label{eq:reflectingBC}
\bar z(t) = z(t).
\ee
(We are suppressing spatial arguments, so this really means ${\bar z(x,t)=z(x,t)}$.)

Together with the domain wall boundary condition on $z(x,0)$, Eq.~\ref{eq:reflectingBC}
for the boundary values of $z(x,\tau)$ and $\bar z(x,\tau)$ at $\tau=t$
is sufficient to fully specify a solution to the saddle point equations.

Note that, if we were to neglect the nonlinear terms in the equations of motion Eqs.~(\ref{eom1}),~(\ref{eom2}), 
then Eq.~(\ref{eq:reflectingBC}) would convert the problem into a diffusion process of duration $2t$ for a field
$\phi(\tau)$, with ${\phi(\tau)=z(\tau)}$ for $0\leq \tau\leq t$, 
and ${\phi(\tau)=\bar z(2t-\tau)}$ for ${t\leq \tau\leq 2t}$.
That is, the field $z(\tau)$ that is diffusing in the ``forward'' time direction is reflected back from the boundary condition as the field $\bar z(t)$ that diffuses as we progress ``backward''.

One important consequence of the boundary condition 
in Eq.~(\ref{eq:reflectingBC})
is that $z$ and $\bar{z}$ can be chosen to be real. If $z(x,\tau=0)$ is real for all $x$, then eqs. (\ref{eom1}), (\ref{eom2})  imply that $z$ and $\bar{z}$ are real everywhere.

We have converted the problem from one with both initial and final boundary conditions to one with just an initial boundary condition. If we fix $z_I$ and then minimize the action $\mathcal{S}$ while allowing $z_F$ to vary, we find that $\mathcal{S}$ is minimized when 
$z_F^*=z(t)$.

When this condition is satisfied, the boundary action (including the contribution from $\langle \tilde{\psi}^+ \vert z(t) \rangle$) becomes
\begin{equation} \label{eq:boundaryafterminimising}
    \mathcal{S}_\mathrm{bdry} = -\frac{1}{2}\int\mathrm{d}x\ln\frac{1+\bar{z}(0)z_I}{1+z_I^* z_I}
\end{equation}
which depends only on the initial configuration $z_I$ (given by $\langle C_A\vert$) and the value of $\bar{z}$ at $\tau=0$.
Once the saddle-point is found, the total action is computed using Eq.~(\ref{continuumaction}) with boundary term Eq.~(\ref{eq:boundaryafterminimising}). This action then gives the saddle-point approximation to $\check S_2$, by Eq.~(\ref{purityequation}) .

\section{Applications of the saddle point approximation}
\label{eq:saddlepointapplications}

The saddle point approximation introduced in the previous section can be used to calculate the purity of arbitrary subsystems as a function of time, in chains with either open or periodic boundary conditions. For concreteness, we will focus on a chain of length $L=L_A+L_B$ with open boundaries, where the subsystem $A$ is the rightmost $L_A$ sites. We will normally fix $L_A=L_B$ for greater symmetry.
This will allow us to study how bipartite entanglement grows over time and eventually saturates,
on a timescale ${L^2/\Delta^2}$, if the system is finite.

\begin{figure}
    \includegraphics[width=\columnwidth]{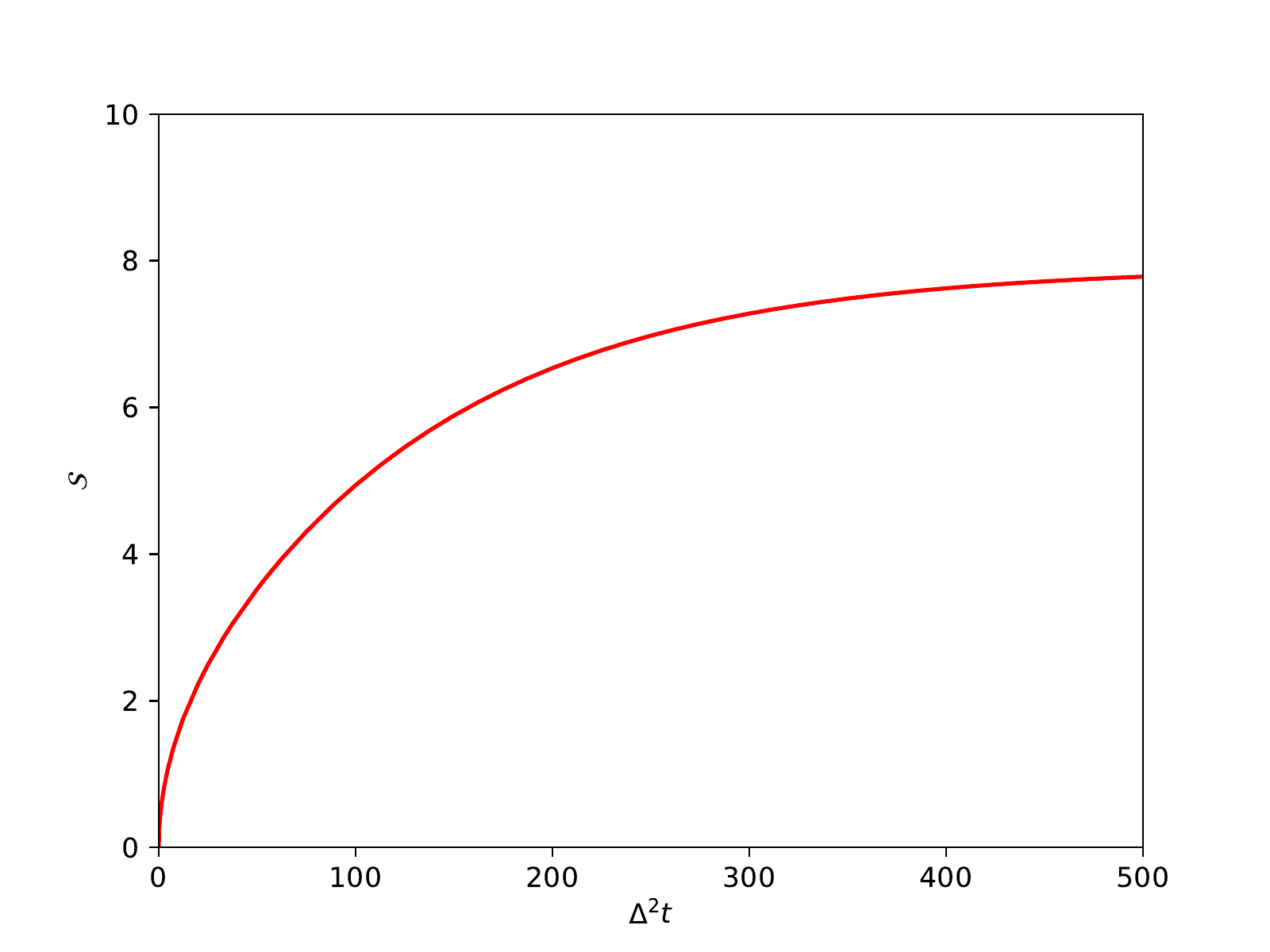}
    \caption{An example of $\mathcal{S}$ against $\Delta^2 t$ for $L_A = L_B = 50$. To find $\mathcal{S}$, the equations of motion are solved numerically, then these solutions are substituted into (\ref{continuumaction}). As expected, $\mathcal{S}$ initially grows rapidly, before eventually approaching a constant.}  \label{figsaddlepoint}
\end{figure}

In Figure \ref{figsaddlepoint} we plot the saddle-point prediction for $\mathcal{S}$ against $\Delta^2 t$ 
for the 
 sharp domain wall
 boundary condition on $z_I(x)$ described above, 
 with
 ${L_A = L_B = 50}$ sites either side of the discontinuity. We calculate $\mathcal{S}$ by numerically solving the  spatially discrete versions of the 
equations of motion (\ref{eom1}) and (\ref{eom2}) and substituting into the action (\ref{continuumaction}).

The discretization we use is the natural one given by the lattice Hamiltonian. To numerically solve the equations, we initially guess $\bar z=0$ everywhere and evolve $z$ forward to generate a guess for $z$. We then evolve $\bar z$ backwards using this guess for $z$, to generate a new guess for $\bar z$, and repeat this procedure until the guesses are consistent.

In this section we will discuss in turn: the equilibrium entanglement of the finite chain and the approach to this equilibrium as ${\Delta^2 t\gg L^2}$; the infinite chain and ${\Delta^2 t\ll L^2}$; the interpolation between these two limits and a post hoc justification for the saddle point approximation. In the next section, we compare the predictions of the saddle point approximation to direct numerical simulations of the Majorana chain.

\subsection{Equilibrium entanglement}
\label{sec:eqmentanglement}

As a sanity check we can use the saddle-point approximation to calculate the averaged purity in an equilibrium ``infinite temperature'' pure state. To do this, we take a finite system of length $L$ and take $t \to \infty$. We take $L$ to be a multiple of four and take the subsystem to be rightmost $L/2$ Marojanas. In the continuum picture, we take our system to lie in the interval $-L/2<x<L/2$ and our subsystem to lie in the interval $0<x<L/2$. The initial boundary condition is that $z_I(x)$ is a step function with azimuthal angle $\theta=0$ for $x<0$ and $\theta=\frac{\pi}{2}$ for $x>0$ (where the azimuthal angle is defined by $z=e^{-i\phi}\cot\frac{\theta}{2}$).

For $t \to \infty$, $z$ and $\bar{z}$ relax to a trivial steady state as $\tau\to\infty$. The only steady state consistent with the boundary conditions (including the spatial boundaries)  is one in which $z=\bar{z}$ and they are spatially uniform, that is, they both approach some constant value. We can always choose a basis such that this constant is zero by rotating around the $y$-axis (this will leave $z$ and $\bar{z}$ real because real $z$ corresponds to the $xz$-plane). In the symmetrical case $L_A=L/2$, we rotate by $\frac{3\pi}{4}$ so that our initial domain wall has $\theta=\frac{3\pi}{4}$ for $x<0$ and $\theta=\frac{5\pi}{4}$ for $x>0$. This means taking $z_I(x)= \mathrm{sgn}(x) c$ where $c \equiv \tan\frac{\pi}{8} = \sqrt{2}-1$. If $\bar{z}(t)=0$, then the equation of motion (\ref{eom2}) means that $\bar{z}$ is also zero at all earlier times.

Switching to a basis where $\bar{z}=0$ everywhere means that both the Berry phase term and the Hamiltonian term in eq. (\ref{continuumaction}) are zero. The action is determined solely by the boundary term (\ref{eq:boundaryafterminimising}). Substituting in $\bar{z}(0)=0$ and $z_I^*z_I = c^2$ gives
\begin{equation}
    \mathcal{S}_\mathrm{eq} = \frac{L}{2}\ln(1+c^2)
\end{equation}
The saddle-point prediction equilibrium purity is therefore
\begin{equation} \label{saddlepointequilibrium}
    \overline{\Tr[\rho_A^2]} \approx K^\infty_{\mathrm{fl}} (1+c^2)^{-L/2}
\end{equation}
where $ K^\infty_{\mathrm{fl}}$ comes from integrating over fluctuations around the stationary path (note that these quantum fluctuations are not directly related to the fluctuations in $S_2$ over different realizations of the noise $\eta$). The predicted purity is exponentially small in $L$ as expected.

In the Appendix \ref{apppurityincludingfluctuations}, we calculate the equilibrium purity in the thermodynamic limit more carefully for $L_A=L/2$ by including fluctuations around the saddle point solution. We find that the equilibrium purity approaches
\begin{equation}
    \overline{\Tr[\rho_A^2]} = (1+c^2)^{-(L+1)/2}
\end{equation}
corresponding to $K^\infty_{\mathrm{fl}} = (1+c^2)^{-1/2}$, so $K_{\mathrm{fl}}$ remains $O(1)$ as $L \to \infty$.

We can generalize the saddle-point prediction eq. (\ref{saddlepointequilibrium}) to arbitrary subsystem size $L_A$ (and $L_B=L-L_A$)
\begin{equation}
    \mathcal{S}_\mathrm{eq} = \frac{L_A}{2}\ln(1+c_{-}^2) + \frac{L_B}{2}\ln(1+c_{+}^2)
\end{equation}
where $c_\pm^2 = (3-q^2-2\sqrt{2-q^2})/(1\mp q)^2$ with ${q=(L_A-L_B)/L}$.

\subsection{Approach to equilibrium}\label{sectionapproachtoeq}

Having computed the steady state entanglement, we consider the approach to it.

In the limit $t \to \infty$, we found $\bar{z}\to 0$ 
(in the natural  basis).
We should expect that for large but finite $t$, $\bar{z}$ will be small but non-zero, and that we can expand the action (\ref{continuumaction}) in powers of $\bar{z}$.
To leading order in $\bar{z}$ this gives
\begin{align} \label{smallzbaraction}
&\mathcal{S}   \approx
    -\frac{1}{2}\int \dd \tau\,\dd x\,[\dot{\bar{z}}z-\bar{z}\dot{z}]
    +4\Delta^2\int \dd \tau\,\dd x\,\bar{z}'z' + \mathcal{S}_\mathrm{bdry},
    \\
&    \mathcal{S}_\mathrm{bdry}  = -\frac{1}{2}\int \dd x\,\mathrm{ln}
    \frac{1+\bar{z}(\tau=0)z_I}{1+z^*_I z_I}
.\end{align}   
If we vary this action with respect to $z$ and $\bar{z}$ and set it to zero, we get the following diffusion equations:
\begin{align}
    \dot{z} &= + 4\Delta^2 z'', \\
    \dot{\bar{z}} &= - 4\Delta^2 \bar{z}''.
\end{align}
As discussed around Eq.~(\ref{eq:reflectingBC}), 
for  late times ${\Delta^2 t \gg 1}$ we  take the boundary condition at $\tau=t$ to be ${z(t) = \bar{z}(t)}$.
Therefore,
$z$ simply ``reflects'' off this boundary as $\bar{z}$: 
The initial domain wall $z(x,0)$ diffuses forwards for time $t$ as $z(x,\tau)$, then backwards for a further time $t$ as $\bar{z}(x,\tau)$.
That is, diffusion for a total time $2t$ transforms $z(x,0)$ into  $\bar{z}(x,0)$.

Assuming these equations of motion hold, we can substitute them into the action (\ref{smallzbaraction}). Integrating by parts, the Berry phase term perfectly cancels the Hamiltonian term, so the action is once again determined solely by the boundary terms.

By solving the diffusion equation exactly for the sharp domain wall, keeping only the slowest mode, and substituting into eq. (\ref{smallzbaraction}), we get
\begin{equation} \label{smallzbaraction2}
    \mathcal{S} \approx
    \mathcal{S}_{\mathrm{eq}}
    -\frac{4Lc^2}{\pi^2}\exp\left(-8\pi^2\frac{\Delta^2 t}{L^2}\right)
\end{equation}
where ${\mathcal{S}_{\mathrm{eq}} = \frac{L}{2}\ln(1+c^2)}$ and we have used the fact that $\bar{z}$ is small to approximate the boundary term ${\ln(1+\bar{z}(0)z(0))\approx \bar{z}(0)z(0)}$. The terms we have dropped are of order $\bar{z}^2$ and are therefore exponentially smaller in $t$ than the last term in eq. (\ref{smallzbaraction2}).

\begin{figure}
    \includegraphics[width=\columnwidth]{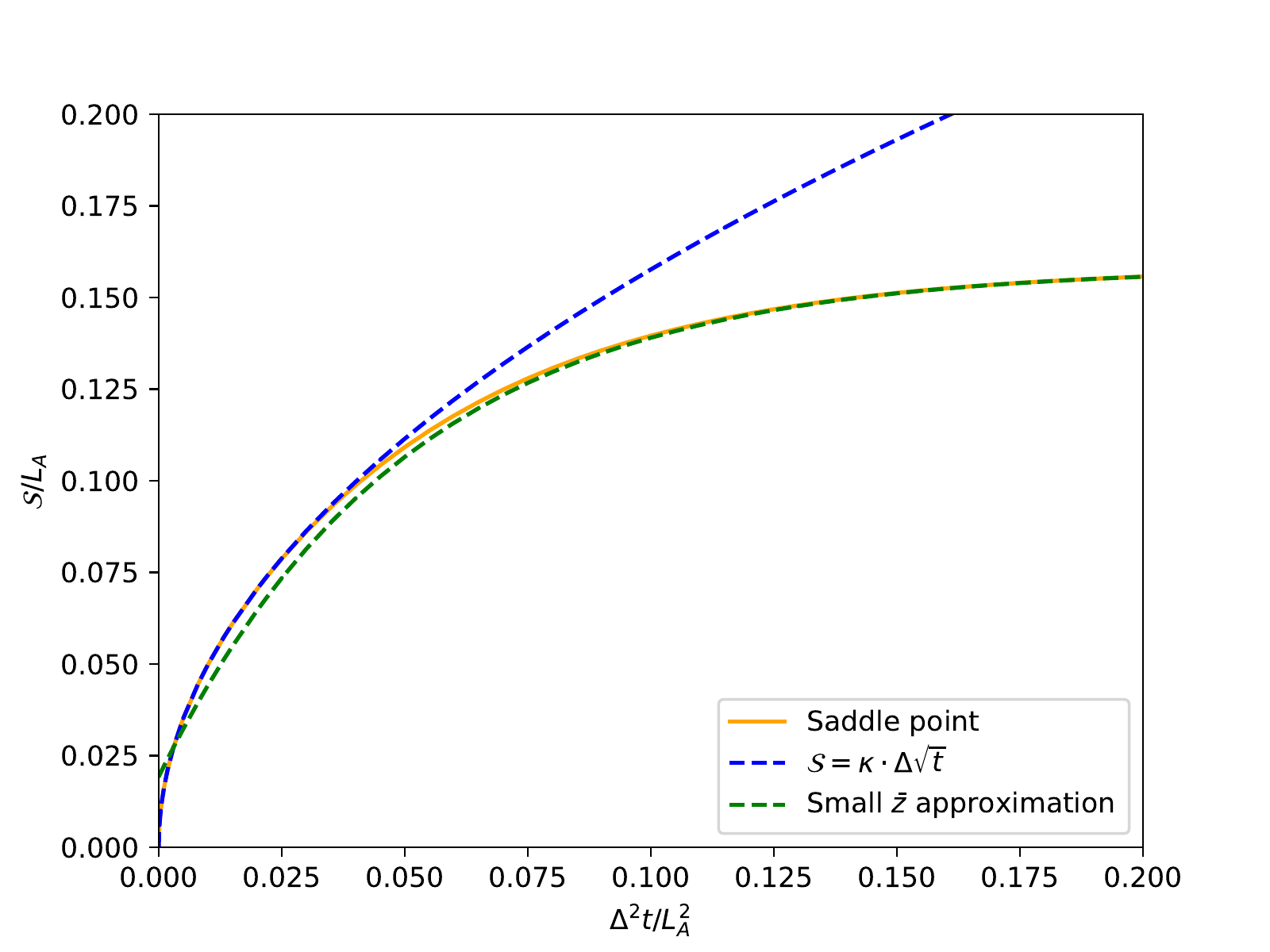}
    \caption{A comparison of the full numerical solution for $\mathcal{S}$ (labelled ``Saddle point'') against the small $\bar{z}$ approximation (\ref{smallzbaraction2}), and the $\sqrt{t}$ growth (\ref{eqearlytimes}) found at early times. Here $L_A=L_B=50$. The full solution approaches the small $\bar{z}$ approximation when $\Delta^2 t$ is comparable to $L^2$. The full solution grows approximately as $\kappa\cdot\Delta\sqrt{t}$ when $\Delta^2 t\ll L^2$.} \label{figsaddlepointapprox}
\end{figure}

In Figure \ref{figsaddlepointapprox}, we compare eq. (\ref{smallzbaraction2}) to numerical solutions of the full equations of motion (\ref{eom1}) and (\ref{eom2}) substituted into the full action (\ref{continuumaction}).

We can understand this exponential approach to $\mathcal{S}_\mathrm{eq}$ by conisdering the energy spectrum of the spin chain. As the imaginary time $t\to\infty$, the approach to the ground state is dominated by the excited state with the lowest energy. The spin chain is a Heisenberg ferromagnet and therefore has quadratic dispersion, meaning it has an energy gap $\delta E \propto 1/L^2$ and this contribution decays as $\exp(-\delta E \cdot t)$ with $\delta E = 8\pi^2\frac{\Delta^2}{L^2}$ (similar to \cite{bernard2022dynamics}).

\subsection{Entanglement growth at early times} \label{sectionearlytimes}

In the two previous subsections, we assumed that the system size $L$ was finite,  and that $\Delta^2 t\gg L^2$ so that correlations had been able to spread throughout much of the system (the entanglement was comparable to the equilibrium entanglement). This allowed us to assume that $\bar{z}$ is small everywhere.

As $L$ becomes larger, it takes indefinitely long to reach this regime, and if $L\to\infty$ then we cannot use this approximation at all. On the other hand, as $L\to\infty$, the length-scale $L$ drops out of the problem, because for $\Delta^2 t\ll L^2$, the physical boundaries of the system are not important. We can use scaling arguments to understand how entanglement grows in this regime.

Let’s focus on the case where the system is an infinite chain and we are interested in the purity of the semi-infinite subsystem $x>0$. We will also assume that $\Delta^2 t \gg 1$ so that we can take the $\tau=t$ boundary condition to be simply $z(t) = \bar{z}(t)$.

In this case, we rewrite the problem in terms of dimensionless coordinates $\tilde{\tau}=\tau/t$ and $\tilde{x}=x/(\Delta\sqrt{t})$. The solutions $z(\tilde{x},\tilde{\tau})$ and $\bar{z}(\tilde{x},\tilde{\tau})$ obey the dimensionless equations of motion
\begin{align}
    \partial_{\tilde{\tau}}z &= + 4 \left( \partial_{\tilde{x}}^2 z - \frac{2\bar{z}(\partial_{\tilde{x}}z)^2}{1+\bar{z}z} \right) \\
    \partial_{\tilde{\tau}}\bar{z} &= - 4 \left( \partial_{\tilde{x}}^2\bar{z} - \frac{2z(\partial_{\tilde{x}}\bar{z})^2}{1+\bar{z}z} \right)
\end{align}
which are independent of $t$. The boundary conditions are $z(\tilde{x},0)=\mathrm{sgn}(\tilde{x})c$ and $z(\tilde{x},1)=\bar{z}(\tilde{x},1)$ which are also independent of $t$. This means that $z(\tilde{x},\tilde{\tau})$ and $\bar{z}(\tilde{x},\tilde{\tau})$ have a single solution in this regime for any $t$.

If we rewrite the action (\ref{continuumaction}) using the dimensionless coordinates $\tilde{\tau}$ and $\tilde{x}$, we therefore end up with something of the form
\begin{equation} \label{eqearlytimes}
    \mathcal{S} = \kappa \cdot \Delta\sqrt{t}
\end{equation}
where $\kappa$ is a numerical constant independent of $t$ and $\Delta$, equivalent to calculating the action with $t=\Delta=1$. We compare (\ref{eqearlytimes}) to the a full numerical solution for $\mathcal{S}$ in Figure \ref{figsaddlepointapprox}. This means that, in the regime $1 \ll \Delta^2 t \ll L^2$, the average purity decays as \footnote{We have neglected time dependence of the $O(1)$ correction due to $K_\mathrm{fl}$.}
\begin{equation} \label{decayofpurity}
    \overline{\Tr[\rho_A^2]} \propto e^{-\kappa\Delta\sqrt{t}},
\end{equation}
that is
\be
\check{S}_2 \sim  \kappa \cdot \Delta \sqrt{t}.
\ee

\begin{figure}
\centering
\includegraphics[width=\columnwidth]{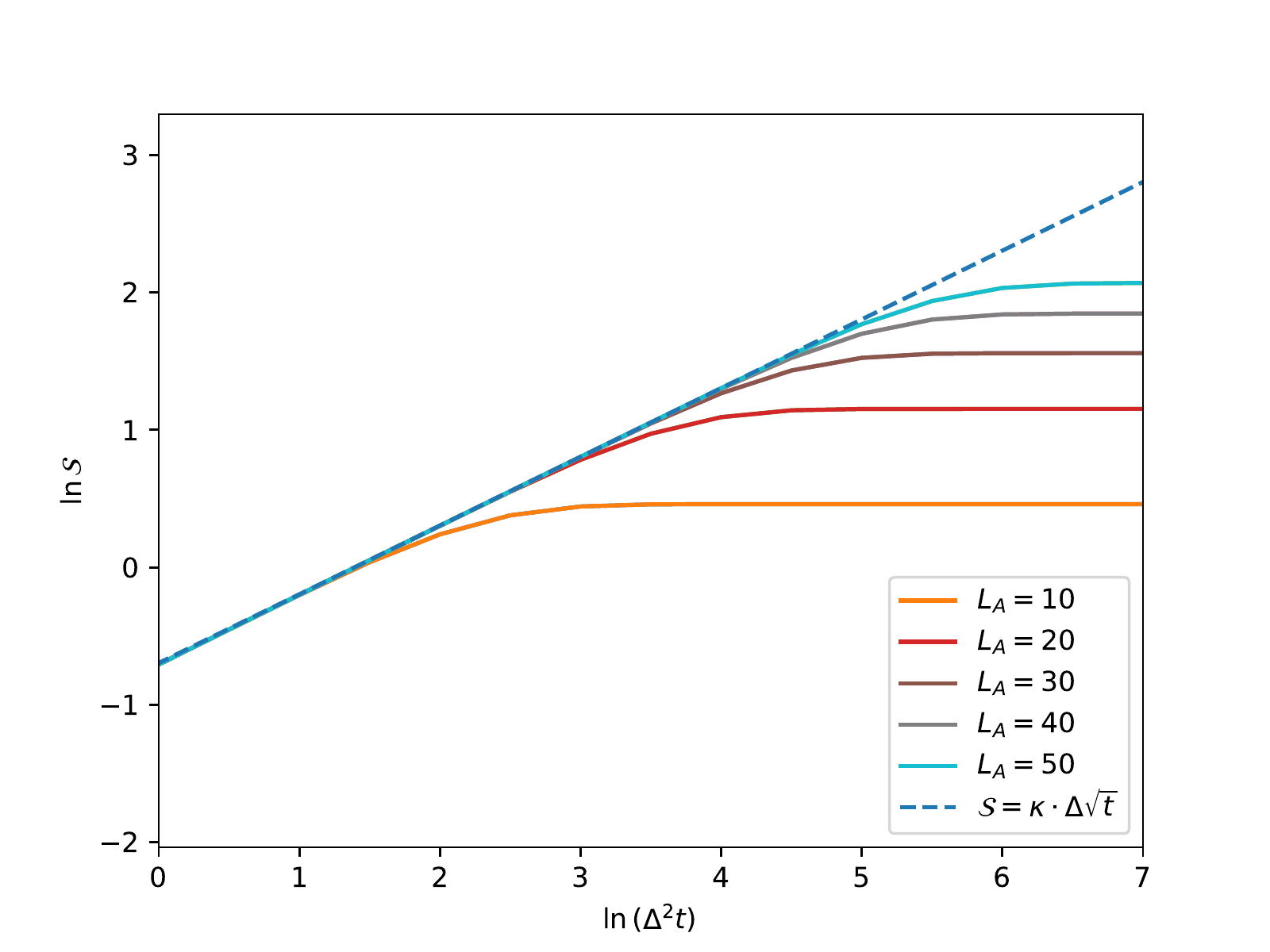}
\caption{Plot of $\ln\mathcal{S}$ against $\ln (\Delta^2 t)$ for $L_A=L_B=10,20,30,40,50$. The asymptotic $\sqrt{t}$ behaviour is shown by the dashed line. The $y$-intercept gives the value of $\ln \kappa$.}\label{fig:loglogplot}
\end{figure}

In Figure \ref{fig:loglogplot} we plot numerical calculations of $\ln\mathcal{S}$ against $\ln (\Delta^2 t)$ for increasing values of $L$ (we take $L_A=L_B=L/2$). We see an approach to the asymptotic form $\ln\mathcal{S} = \ln\kappa + \frac{1}{2}\ln (\Delta^2 t)$.

To connect this with the limit $\Delta^2 t \gg L^2$, we can write a more general scaling form for the saddle-point action $\mathcal{S}$
\begin{equation}
    \mathcal{S} = g\left(\frac{L}{\Delta\sqrt{t}}\right) \Delta\sqrt{t}
\end{equation}
where $g(\alpha) \to \kappa$ as $\alpha \to \infty$ but $g(\alpha)\sim \frac{1}{2}s_\mathrm{eq}\alpha$ as $\alpha\to 0$ (where we have used the equilibrium action density ${s_\mathrm{eq}\equiv\mathcal{S}_\mathrm{eq}/L_A}$). 

In Appendix \ref{app:scalingfunction}, we plot the function $g(\alpha)$. By plotting $g(\alpha)$ for increasing values of $L$ we can approximate $\kappa\approx 0.49855\dots$.

\subsection{Validity of the saddle-point approximation}
\label{sec:saddleptvalidity}
 
The saddle-point approximation assumes that the path integral is dominated by paths close to the classical trajectory.
That is, there should be a large parameter that suppresses deviations from this trajectory. 
In our case, this large parameter is given by the characteristic timescale for the evolution.

Consider for definiteness the regime discussed above where the total time $t$ is large, and $L$ is of order $\sqrt{t}$.
Rewriting the continuum action (\ref{continuumaction}) in terms of the dimensionless coordinates of the previous section, ${\tilde \tau= \tau/t}$ and ${\tilde x = x/(\Delta \sqrt{t})}$, we find that $\Delta \sqrt{t}$ stands outside the action as a large parameter, with all the other quantities that appear in the action generally being of order $g(\alpha)$ (where $\alpha\equiv L/(\Delta\sqrt{t})$). For any fixed value of $\alpha>0$, the limit $\Delta\sqrt{t}\to\infty$ leads to an indefinitely large prefactor in the action, justifying a saddle-point treatment with large parameter $\Delta\sqrt{t}$.

A standard idea for the ferromagnetic chain is that we can understand this as being similar to a large $S$ limit \cite{takahashi1987few}. Heuristically, in the low-energy states that are relevant at large times, large blocks of spins of size ${M\gg 1}$ are very well aligned, so they act as individual spins with~${S = M/2}$.

\section{Numerics}

To test the validity of the saddle-point approximation, we compared the predictions of the approximation with direct simulations of the Trotterized noisy Majorana chain.

To simulate the Majorana chain we don’t need to keep track of the entire many-body wave function (which has $2^{L/2}$ complex coefficients). Because it describes free fermions, we only need to keep track of the time-dependence of the two-point correlation functions
\be\label{eq:Gmatrix}
G_{ij} \equiv \frac{i}{2} \< [\gamma_i^a,\gamma_i^b]
\>,
\ee
(which are $L(L-1)/2$ real numbers) for a given realization of the noise.
Entanglement entropies can be computed from $G$ using standard techniques \cite{Ingo_Peschel_2003,Vidal_2003,Bravyi_2004,calabrese2005evolution,Hackl_2021}. This approach allows much larger system sizes.

\begin{figure}
\centering
\includegraphics[width=\columnwidth]{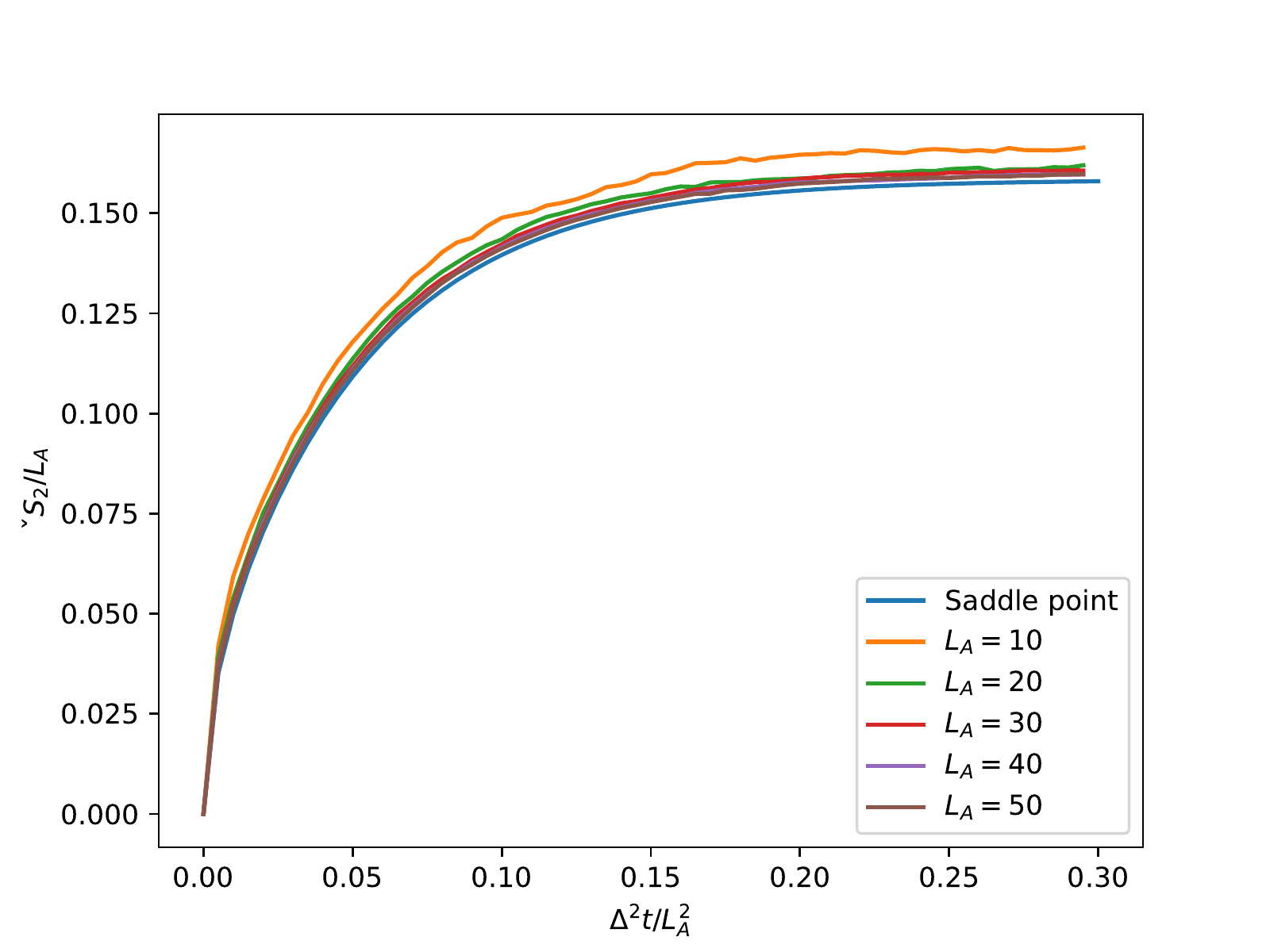}
\caption{Plots of $\check{S}_2/L_A$ against $\Delta^2 t/L_A^2$ for $L_A=10,20,30,40,50$ from direct simulations of the Majorana chain, compared with the saddle-point prediction.  The averages are calculated over 500 random realizations of the random unitary circuit with $\Delta=0.1$. For finite $L_A$, we  expect $\check{S}_2/L_A$
to match to the saddle point prediction up to a correction  of order $1/L_A$ (see Fig.~\ref{fig:convergence}).} \label{fig:sims}
\end{figure}

Figure \ref{fig:sims} shows $\check{S}_2/L_A$ plotted against $\Delta^2 t/L_A^2$ for increasing values of $L_A$. The average was taken over 500 trials. These simulations are consistent with a collapse to the saddle-point prediction.

\begin{figure}
\centering
\includegraphics[width=\columnwidth]{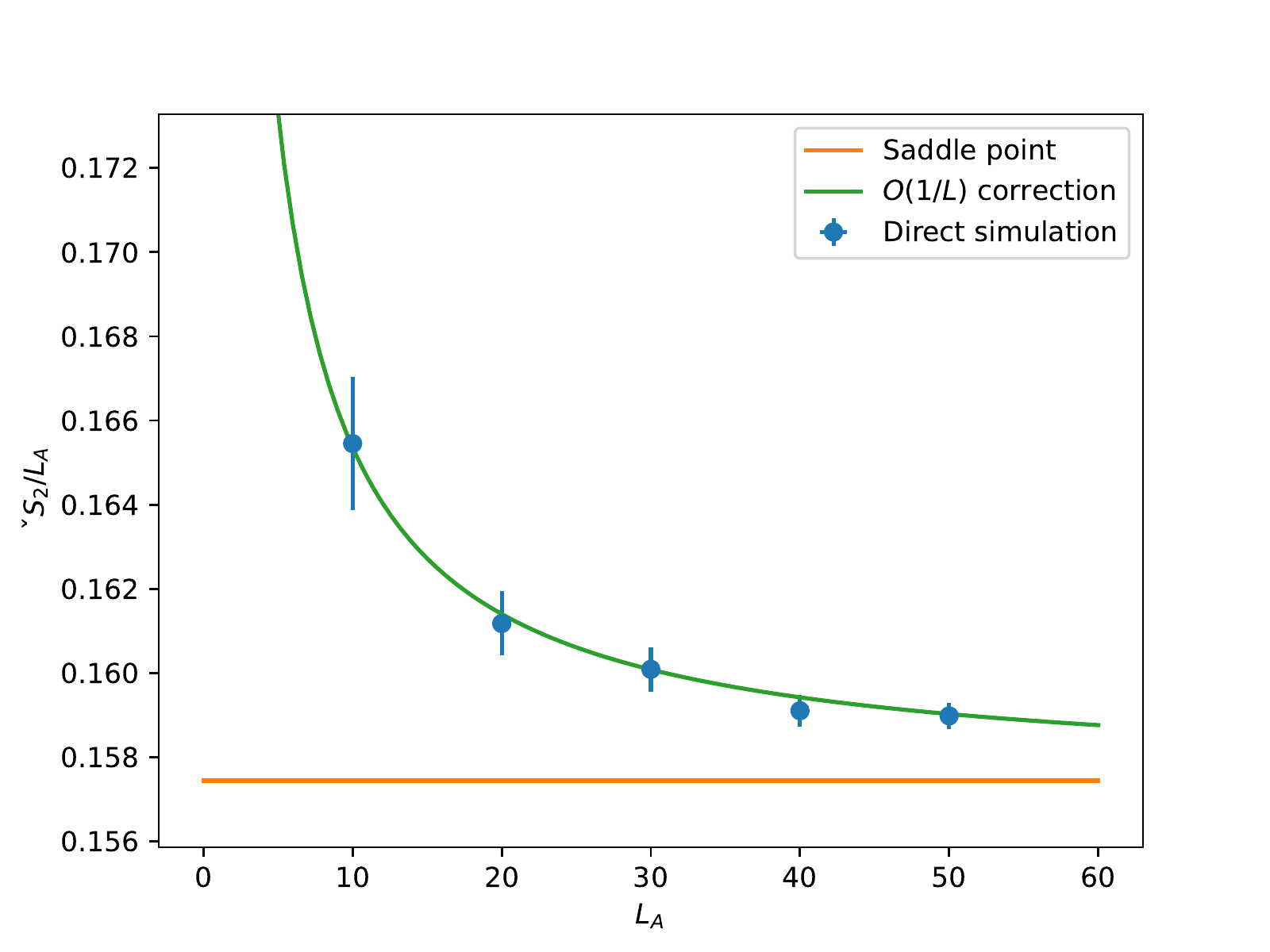}
\caption{Values of $\check{S}_2/L_A$ against $L_A$ for $\Delta^2 t/L_A^2=0.25$. The values were found by averaging over 500 trials, and the error bars show the standard error. The $O(1/L)$ correction to the saddle point prediction for $\check{S}_2/L_A$ is approximated as $-\ln K^\infty_\mathrm{fl}/L$, where $K^\infty_\mathrm{fl}$ is the pre-exponential factor $K_\mathrm{fl}$ in the limit $t\to\infty$.}\label{fig:convergence}
\end{figure}

From the discussion of the effective large parameter in the saddle-point treatment in  Sec.~\ref{sec:saddleptvalidity},
corrections to $\check S_2$ from fluctuations around the saddle point are of 
relative order $1/L$
compared to the saddle-point result, 
for any fixed value of the scaling variable  ${\Delta^2 t/L_A^2}$.
See for example the result 
${\check{S}_2=\frac{L+1}{2}\ln(1+c^2)}$  for the late-time entanglement in Sec.~\ref{sec:eqmentanglement}.
To test this, Fig.~\ref{fig:convergence} shows the result for $\check{S}_2/L_A$ 
found from simulations, 
plotted against $L_A$,
for a fixed value of the scaling variable ${\Delta^2 t/L_A^2=1/4}$.
We see that the dependence on $L_A$ is indeed consistent with a $1/L$ convergence to the saddle point prediction as $L\to\infty$.

\subsection{Small size of statistical fluctuations}
\label{sec:fluctuations}

\begin{figure}
    \centering
    \includegraphics[width=\columnwidth]{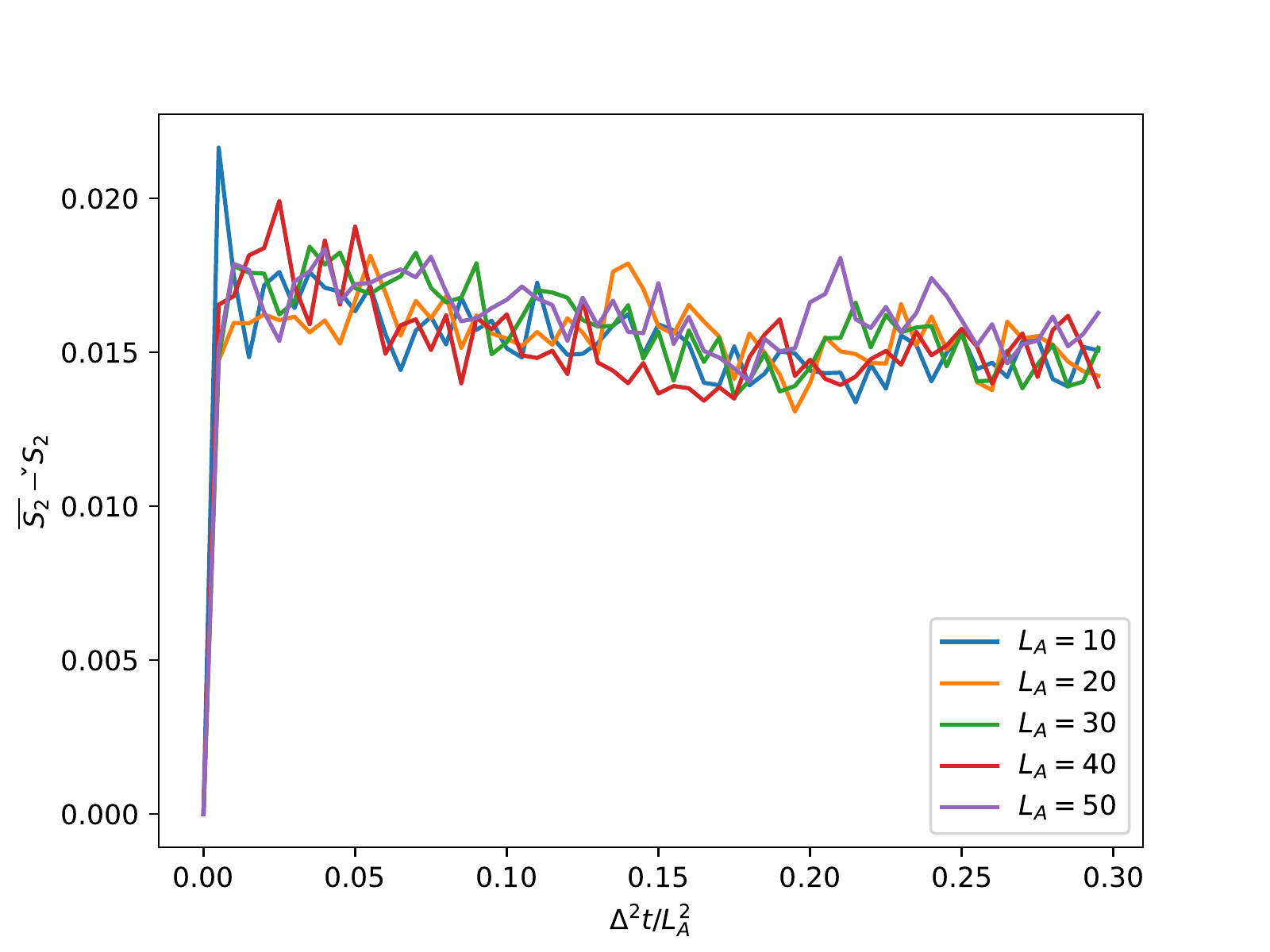}
    \caption{Plot of $\overline{S_2}-\check{S}_2$ against $\Delta^2 t/L_A^2$ for $L_A=L_B=10,20,30,40,50$. The mean second R\'enyi entropy $\overline{S_2}$ is larger than $\check{S}_2=-\ln\overline{\Tr[\rho_A^2]}$ given that $e^{-\overline{S_2}}\leq\overline{e^{-S_2}}$, but the difference does not appear to grow with $L_A$.
    This suggests that $\overline{S_2}$ and $\check{S}_2$ are equal to leading order in $t$. The data was generated by simulating the Majorana chain and averaging over 500 trials. 
    }\label{fig:fluctuations}
\end{figure}

Next we examine the size of statistical fluctuations of $S_2$ 
arising from the noise in the physical Hamiltonian.
Empirically, we find these are surprisingly small, such that $\check S_2$ as defined by Eq.~(\ref{eq:definecheckS})
coincides at leading order with $\overline{S_2}$.

Fig.~\ref{fig:fluctuations} shows difference between these two kinds of average, as a function of the scaling variable $\Delta^2 t/L_A$, for various system sizes (with $L=2L_A$).
Notably, this figure strongly suggests that this difference remains bounded as $L\rightarrow\infty$.
(This is in contrast to interacting random circuits, where the two quantities have slightly different growth rates \cite{ZhouNahum}.)

The difference between the two quantities may expressed using the cumulant expansion:
\ba
e^{-\check S_2}
& = \overline{e^{- S_2}} 
 = \exp \lf \sum_{k=1}^\infty \f{(-1)^k}{k!} \,  \overline{S_2^k}^c
\ri,
\end{align}
where $\overline{S_2^k}^c$
denotes the $k$th cumulant.
Therefore the fact that the difference between $\check S_2$ and $\overline{S_2}$ is order 1 at large size
requires that 
the cumulants with $k>1$ are at most of order 1 size.

As mentioned in Sec.~\ref{sec:stochastic}, 
the distribution of the pure state at asymptotically late times ($t\rightarrow\infty$ for fixed $L$) 
is invariant under ${\rm SO}(L)$ transformations \cite{bernard2021entanglement,
bianchi2021page,
bianchi2022volume,
nadal2010phase,
nadal2011statistical}. 
This means that  the ${L\times L}$ correlation matrix $G_{ij}$ (Eq.~(\ref{eq:Gmatrix})) 
is given by
\ba
G & = O \Sigma O^T,
&
\Sigma = 
\lf
\begin{array}{cc}
   0  & 1 \\
  -1   & 0
\end{array}
\ri
\oplus \ldots \oplus
\lf
\begin{array}{cc}
   0  & 1 \\
  -1   & 0
\end{array}
\ri,
\end{align}
where $O$ is a Haar-random 
${\rm SO}(L)$ matrix, and $\Sigma$ is the correlation matrix for the initial state. 
For the case of fermions with a conserved ${\rm U}(1)$ charge, 
the analogous expression has been used to compute the probability distribution of the entanglement in the late-time steady state using random matrix techniques,  and the cumulants (at large time) have the scaling \cite{bernard2021entanglement} 
\be
\overline{S_2^k}^c \sim L^{2-k}
\ee
(the variance was also considered in \cite{bianchi2022volume}).
It is expected that the same scaling will generalize to the Majorana case due to the link between the invariant measure and random matrix theory \cite{bernarddelucapiroliunpublished}. If so, it is consistent with Fig.~\ref{fig:fluctuations}, with the fluctuations (from the second cumulant) being of order 1.

The result in this Section is perhaps surprising: it means that for these free-fermion (unitary \cite{sigmamodelmeasurement}) dynamics we obtain asymptotically exact results for the second R\'enyi entropy $S_2$ 
in a typical realization without needing taking a replica limit, again in contrast to random interacting circuits.

\section{Effect of interactions}

So far, we have been considering the free-fermion model given by eq. (\ref{majoranahamiltonian}). We have seen that this leads us to a Heisenberg model with a continuous ${\rm SO}(2N)$ symmetry, in stark contrast with the discrete permutation symmetry which emerges from interacting models. We might ask, how does adding weak interactions to our free-fermion model qualitatively change the emergent Heisenberg model?

\subsection{An interacting model}

Let's focus on a particular weakly interacting model. The simplest terms we can add to our Hamiltonian are terms involving four Majorana operators, so let's consider adding a term to the Hamiltonian of the form
\begin{equation}\label{eq:interactionH}
     H_\gamma^\mathrm{int}(t) = - \sum_i \eta_i'(t) \gamma_i \gamma_{i+1} \gamma_{i+2} \gamma_{i+3}
\end{equation}
where $\eta_i'(t)$ is Gaussian noise with zero mean and $\langle \eta_i'(t)\eta_j'(t')\rangle = \Delta_I^2 \delta_{ij} \delta(t-t')$. This is equivalent to adding gates to the  unitary circuit of the form
\begin{equation}
    u_I(\eta') = \exp(i \eta' \gamma_i \gamma_{i+1} \gamma_{i+2} \gamma_{i+3})
\end{equation}
where $\eta'$ is a Gaussian random variable with zero mean and variance $\Delta_I^2\delta t$. In order to study weak interactions, we will assume $\Delta_I^2 \ll \Delta^2$.

When we perform the mapping from $2N$-replicas of this model to a Heisenberg model, this gives us noise-averaged replicated gates of the form
\begin{equation}
    \overline{\tilde{u}_I} = \exp\left(-\frac{\Delta_I^2\delta t}{2}
    \left[2N+2\sum_{a<b}^{2N}(-1)^{a+b}A_i^{ab}A_{i+1}^{ab}A_{i+2}^{ab}A_{i+3}^{ab}\right]\right)
\end{equation}
which, in the limit $\delta t \to 0$, is equivalent to adding an interaction Hamiltonian $H_I$ to our original Hamiltonian given by
\begin{equation} \label{interactionhamiltonian}
    H_I = \Delta_I^2\sum_i
    \left[N+\sum_{a<b}^{2N}(-1)^{a+b}A_i^{ab}A_{i+1}^{ab}A_{i+2}^{ab}A_{i+3}^{ab}\right]
\end{equation}
This interaction Hamiltonian explicitly breaks the continuous ${\rm SO}(2N)$ symmetry of the original model down to a discrete symmetry. 

Interacting random unitary circuits generically have an $N$-permutation symmetry in the replica formalism, leading to $N!$ distinct ground states. However, the interacting model introduced above has additional symmetries, which we discuss in the next section. These additional symmetries lead a spin model with $2^{N-1}N!$ relevant ground states.

\subsection{Symmetries}
\label{sec:symmetriesgeneral}

We can understand heuristically why $2N$-replicated generic interacting models have $N!$ ground states by considering the path integral formulation. When summing over trajectories in the $2N$-replicated system, we are summing over $N$ forward trajectories and $N$ backward trajectories. Each forward trajectory contributes a factor of ``$e^{i\mathcal{S}}$'' (where $\mathcal{S}$ is the action of the trajectory) and each backward trajectory contributes a factor of ``$e^{-i\mathcal{S}}$''.\footnote{More precisely, in the discrete circuit ``$e^{iS}$'' becomes a product of local unitary amplitudes.} After averaging over noise, the sets of trajectories which dominate at late times are ones in which every forward trajectory is paired with a backward trajectory, so that the phases cancel \cite{Zhou_2020}. There are $N!$ different ways of doing this, corresponding to the $N!$ distinct ground states of the emergent ferromagnet.

However, the interacting model introduced in the previous section is not a generic interacting model. For one thing, it is a fermion model, which means it has a global $\mathbb{Z}_2$ symmetry generated by the fermion parity operator $U_p = (-i)^{L/2}\prod_i \gamma_i$.
This symmetry means that for any trajectory with action $\mathcal{S}$, there is a parity-transformed trajectory which also has action $\mathcal{S}$. When pairing forward and backward trajectories, we can either pair any trajectory with itself or with its parity-transformed version. This gives us $2^N N!$ possible pairings, instead of $N!$.

This may be made  clearer in the language of states. First, let us recall why, for a generic  noisy interacting Hamiltonian  $H_\gamma (t)$, the corresponding replica Hamiltonian has  $N!$ paired ground states.
We start with the identity
\be\label{eq:basicpairingidentity}
  e^{-i \int \dd t H_\gamma (t)}\otimes
 (e^{-i \int \dd t  H_\gamma (t)})^* 
\ket{\mathbb{1}}  =\ket{\mathbb{1}}
\ee
for a single forward and a single backward replica, where 
 $\ket{\mathbb{1}}$ is the paired state 
between the replicas (Sec.~\ref{sssymmetrygroup}).
 This equation  is a restatement of the unitarity of the time-evolution operator prior to  averaging over the noise. 
Eq.~(\ref{eq:basicpairingidentity}) generalizes to a larger number of replicas:
for any pairing $\sigma$ of the forward with the backward layers, we can form the corresponding paired state $\ket{\sigma}$ by taking  the tensor product of  $\ket{\mathbb{1}}$ states for each pair, and 
\be
\left[   e^{-i \int \dd t H_\gamma(t)}\otimes\ldots \otimes
 (e^{-i \int \dd t H_\gamma(t)})^* \right]
\ket{\sigma}  =\ket{\sigma}.
\ee
Averaging over disorder shows that $\ket{\sigma}$ is a ground state\footnote{ In a generic model the replica Hamiltonian could have complex eigenvalues. However unitarity prior to averaging implies that the real part of the eigenvalue is non-negative. It is equal to zero for the ground states.}  of the  replica Hamiltonian $H$,
\be
e^{- \int \dd t H } \ket{\sigma}  =\ket{\sigma}.
\ee

In the presence of a $\mathbb{Z}_2$ symmetry generated by a unitary $U_{\mathbb{Z}_2}$ that commutes with the physical Hamiltonian,  
then the state $\ket{\mathbb{1}_-}$,
\be
\ket{\mathbb{1}_-} \equiv \left( \mathbb{1}\otimes U_{\mathbb{Z}_2}\right)  \ket{\mathbb{1}},
\ee
is also invariant under ${e^{-i \int \dd t H_\gamma (t)}\otimes
 (e^{-i \int\dd t  H_\gamma (t)})^*}$ (cf. Eq.~(\ref{eq:basicpairingidentity})).
After replication, this allows us to form $2^N$ ground states for a given pairing pattern, since for each pair we can use $\ket{\mathbb{1}}$ or $\ket{\mathbb{1}_-}$.

A second property of our model is a ``statistical gauge symmetry''
\be\label{eq:statsymm}
\gamma_i \rightarrow - \gamma_i
\ee
that holds for each site. This is not a symmetry of any individual realization of the noisy Hamiltonian $H_\gamma(t)$, 
but the fact that the probability distribution of the noise is invariant under $\eta_i(t) \rightarrow - \eta_i(t)$ for each  bond means that (\ref{eq:statsymm}) is a local symmetry of the replica Hamiltonian. In other words,  the replica Hamiltonian contains only even numbers of Majorana operators on each individual lattice site.

 This leads to the conversation of chirality $\chi_i = (-i)^N\prod_{a=1}^{2N}\gamma_i^a$ on every site $i$. The boundary conditions we are interested in have $\chi_i = +1$ everywhere, so only ground states with positive chirality are relevant to us. Of the $2^N N!$ ground states, half of them have positive chirality (the ones with an even number of $\mathbb{Z}_2$-transformed pairings), meaning there are $2^{N-1}N!$ relevant ground states.
Below we will see this more explicitly for the case $N=2$, where there are 4 such states.

There is a third property, which is true for the free-fermion Hamiltonian, but which is lost for the interacting Hamiltonian given above. This is that changing the sign of Majorana operators on even sites $\gamma_{2j} \to -\gamma_{2j}$ gives minus its complex conjugate,  i.e.
\be\label{eq:signchangeU}
U_S H_\gamma U_S^\dag =  - H_\gamma^* + \text{const.}
\ee
where $U_S$ implements this sign change.
This property allowed us to make the sign change to get Eq.~(\ref{enlargedsymmetry}), revealing the larger ${\rm SO}(2N)$ symmetry, as opposed to just ${\rm SO}(N) \times {\rm SO}(N)$. This kind of sign change is always possible for free-fermion systems (we can choose an ordering of the Majorana operators $\gamma_i$ such that $i\gamma_i\gamma_{i+1}$ is always real, then let $U_S$ change of the sign of $\gamma_i$ when $i$ is even; then $i\gamma_i\gamma_{i+n}$ is real for odd $n$ and changes sign under $U_S$, and imaginary for even $n$ but doesn't change sign under $U_S$).

In a general interacting system there is no corresponding symmetry. However,
 if instead of the interaction in Eq.~(\ref{eq:interactionH}) we pick a six-Majorana interaction, for example  
 \be
 H_\gamma^6(t) = i\sum_i \eta'_i(t) \gamma_i \gamma_{i+1} \gamma_{i+2} \gamma_{i+3} \gamma_{i+4} \gamma_{i+5},
 \ee
then the property (\ref{eq:signchangeU}) is preserved.
In this case, the interactions  still break the continuous replica symmetry down to a discrete one, but a larger discrete symmetry  than in the generic case.

This corresponds to fact that it is possible to pair forward trajectories with forward trajectories and backward trajectories with backward trajectories. Algebraically, this is because, if we define a paired state between two forward replicas of the form
\be
\ket{S} = \lf \mathbb{1} \otimes U_S \ri \ket{\mathbb{1}},
\ee
then this is invariant under 
  ${e^{-i \int\dd t H_\gamma (t)}\otimes
 e^{-i \int\dd t  H_\gamma (t)}}$. In models of this type, the number of ways of pairing replicas  is $(2N)!/(2^N N!)$ which, when combined with $2^N$-fold degeneracy from the fermionic $\mathbb{Z}_2$ symmetry results in $(2N)!/N!$ paired states, and there is an enlarged $S_{2N}$ permutation symmetry of the replicas.
 
 This property holds for any model where the interaction terms only involve $4n+2$ distinct Majorana operators for integer $n \geq 0$ (and we can sum terms with different values of $n$). This symmetry is therefore broken by any four-Majorana interaction, including (\ref{eq:interactionH})\footnote{If we had \emph{only} these four-Majorana terms, we could recover the symmetry by doing a different sign change, flipping every fourth Majorana operator on even replicas.}.

\subsection{Crossover in entanglement scaling due to interactions}

To make the discussion of interactions more concrete, let's focus on the dynamics of the averaged purity by fixing $N=2$. The chirality on each site $\chi_i$ is still conserved so, like in the free-fermion case, we can restrict the states of the spin chain to states with $\chi_i=+1$ everywhere, leaving a single spin-$\frac{1}{2}$ degree of freedom at each site.

The interaction Hamiltonian (\ref{interactionhamiltonian}) becomes
\begin{align}
    H_I = 2\Delta_I^2 \sum_i
    \left(
    1
 - \Sigma^x_i +  \Sigma^y_i - \Sigma^z_i  \right),
\end{align}
where 
\be
\Sigma^\alpha_i = 
\sigma_i^\alpha \sigma_{i+1}^\alpha \sigma_{i+2}^\alpha \sigma_{i+3}^\alpha.
\ee
This interaction breaks the continuous ${\rm SO}(3)$ symmetry of the free-fermion case down to the dihedral group $D_4$. Instead of a continuum of ground states as $L\to\infty$, only 4 survive, namely uniform product states of $\mid\uparrow\rangle$,$\mid\rightarrow\rangle$,$\mid\downarrow\rangle$ or $\mid\leftarrow\rangle$. 

To compute the purity we must  start with the sharp domain-wall state $\langle C_A \vert = \langle\uparrow\cdots\uparrow\rightarrow\cdots\rightarrow\mid$ and evolve in imaginary time using the Hamiltonian ${H = H_\mathrm{free} + H_I}$. 
We will again address this using the semiclassical equations of motion.
Given that $H_\mathrm{free}$ punishes sudden  changes in the spin direction, at short times the domain wall in $z(t)$ will  relax 
in a way similar to the free-fermion case,   becoming wider and smoother at larger timescale.
However, $H_I$ is an anisotropy term in spin space which penalizes spin directions away from the $x$ and $z$ axes, 
and as a result the domain wall cannot relax to become arbitrarily wide. 
Instead, the domain wall will relax only to a finite width (computed below), 
as a compromise between minimising the contribution to the action from spin gradients  and from the anisotropy term.
This is in analogy to magnets with weakly broken continuous symmetry.

This domain wall width is a crossover lengthscale associated with a renormalization group flow from the universality class of the Gaussian problem to that of the interacting one  (there is also an associated crossover timescale).

The finite domain wall width also implies that the action cost becomes \textit{extensive} in time $t$, rather than scaling as $\sqrt{t}$ as in the free case. 
This implies ``ballistic'' spreading of information:
recall that the linear growth of entanglement entropy implies that correlations exist over a ballistically growing distance, of order $t$, at late times.\footnote{For example, the mutual information $I$ between the regions $A=[-\infty,0]$ and $B=[x,\infty]$ is given in a pure state by $I_x=S_A + S_B - S_C$, where $C=[0,x]$. We have $S_A=S_B\sim \Gamma t$ (where $\Gamma$ is computed below) while $S_C$  is straightforwardly bounded by ${S_C\leq \f{x}{2} \ln 2}$ for a Majorana chain. Therefore 
${I_x\geq \f{\ln 2}{2} (vt - x)}$,
where ${v = 4\Gamma/\ln 2}$. This shows that correlations between $A$ and $B$ are significant if $x$ is within the ballistic lightcone $x\leq vt$.}
This ballistic spreading of correlations should not be confused with ballistic transport of quasiparticles (there is no notion of a quasiparticle in the interacting system). 
Similarly it should be borne in mind that these correlations need not  be detectable by two-point functions of simple operators.

To analyze the finite-width domain wall in more detail, let us consider the limit ${\Delta_I^2 \ll \Delta^2}$. 
In this limit, the steady state is nearly polarized everywhere, although the spin direction $\hat n(x)$ 
varies slowly with position (i.e. the state is approximately a coherent state with spatial derivatives $\ll 1$). In this limit we may continue to use a continuum approximation. 

The equations of motion take the general form
\ba\label{eq:eqmotgeneric}
\dot z & = - (1+ \bar z z)^2 \f{\delta H }{\delta \bar z}, &
\dot {\bar z} & = (1+ \bar z z)^2 \f{\delta H }{\delta z}.
\end{align}
The difference between the free case and the  interacting case is that in the latter these equations permit a non-trivial steady-state solution. As we will show below, there exist solutions
(satisfying the required boundary conditions as ${x\rightarrow \pm \infty}$)
with ${\dot z = \dot{\bar z} = 0}$.
Unlike the complexified solutions that we required in previous sections, these solutions  have 
${\bar z(x)= z^*(x)}$ 
(so $\bar z$ is simply the complex conjugate of $z$). 
This means that the spin polarization vector
\be
\hat n(x) = \frac{1}{1+\bar{z}z} \left( z+\bar{z}, -i(z-\bar{z}), 1-\bar{z}z \right)
\ee  
is simply a  real unit vector, 
and the solution can be interpreted straightforwardly as a domain wall in this polarization.
For a time-independent solution with $z^*=\bar z$, Eqs.~(\ref{eq:eqmotgeneric}) give simply
\be
\f{\delta H[\hat n]}{\delta \hat n(x)} = 0.
\ee
In the limit ${\Delta_I \ll \Delta}$ we may approximate
\begin{equation}
    \Sigma^\alpha_i \approx 
\left[ n^\alpha(x) \right]^4 + \cdots.
\end{equation}
That is, the growing lengthscale as $\Delta_I/\Delta\to 0$
(quantified below) means that 
we can ignore spatial derivatives --- they are an indefinitely small contribution to the already small $H_I$.

Let us now find the coherent state which minimizes the Hamiltonian $H$. Let us parameterize the coherent state using the angle $\theta(x)$ such that the spin direction is $\hat{n}(x) = (\sin\theta(x),0,\cos\theta(x))$. Substituting this into the Hamiltonian $H$ gives a continuum Hamiltonian
\begin{align}
    H &\approx \int \dd x \left[ \Delta^2(\partial_x \theta)^2 + 2\Delta_I^2(1-\cos^4\theta-\sin^4\theta) \right] \\
    &=  \int \dd x \left[ \Delta^2(\partial_x \theta)^2 + \Delta_I^2\sin^2 2\theta \right]
\end{align}
Minimising this Hamiltonian with the usual boundary conditions $\theta(x)\to 0$ as $x\to-\infty$ and $\theta(x)\to\pi/2$ as $x\to+\infty$ yields the solution
\begin{equation}
    \theta(x) = \arctan e^{2Kx}
\end{equation}
where $K \equiv \Delta_I/\Delta$ (we have chosen a domain wall at the origin, but any translation of this solution is also a solution).

\begin{figure}
    \centering
\begin{tikzpicture}
    \newcommand{\spin}[3]{\draw[-stealth] (#1-#2,-#3) -- (#1+#2,#3);}
    \spin{0}{0.141}{0.141}
    \spin{0.5}{0.178}{0.091}
    \spin{1}{0.193}{0.051}
    \spin{1.5}{0.198}{0.027}
    \spin{2}{0.2}{0.014}
    \spin{2.5}{0.2}{0.007}
    \spin{3}{0.2}{0.004}
    \spin{0}{0.141}{0.141}
    \spin{-0.5}{0.091}{0.178}
    \spin{-1}{0.051}{0.193}
    \spin{-1.5}{0.027}{0.198}
    \spin{-2}{0.014}{0.2}
    \spin{-2.5}{0.007}{0.2}
    \spin{-3}{0.004}{0.2}
    \draw[stealth-stealth] (-0.75,0.5) -- (0.75,0.5);
    \node at (0,0.75){$l_\mathrm{int}$};
\end{tikzpicture}
    \caption{Cartoon of the spin polarization $\hat{n}(x)$ in the non-trivial steady state. To the left, all spins are pointing up, and to right, all spins are pointing right. The length scale $l_\mathrm{int}$ corresponds to the thickness of the domain wall. (Note that in reality we are assuming $l_\mathrm{int}\gg a$.)}
    \label{fig:steadystate}
\end{figure}
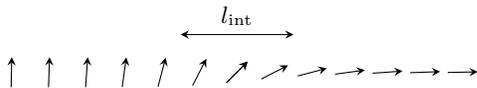

We can immediately infer the length scale of this domain wall (see Fig. \ref{fig:steadystate}) to be ${l_\mathrm{int} \equiv K^{-1}}$, or if we restore the lattice spacing $a$ (previously set equal to 1)
\be
l_\mathrm{int} \equiv \frac{a  \Delta}{\Delta_I}.
\ee

Combining this with the diffusive dynamical scaling on shorter lengthscales (cf. Eq.~(\ref{eom1})) gives the characteristic timescale for relaxation of the domain wall: 
\be
t_\mathrm{crossover} \sim \f{1}{\Delta_I^{2}}.
\ee
A more detailed analysis of the crossover could be obtained by solving the equations of motion
(\ref{eq:eqmotgeneric}) with interaction terms included, to determine how $z$ and $\bar{z}$ approach the above steady-state solution.

The entanglement growth rate 
is given by the bulk action
per unit time, cf. Eq.~(\ref{continuumaction}).
Since the time derivatives vanish, this is simply the value of the Hamiltonian for the domain wall state above, which is  ${H = 2\Delta_I\Delta}$.
This means that at late times $t\gg t_\mathrm{crossover}\equiv\Delta_I^{-2}$, the average purity will decay exponentially in $t$ 
\begin{equation}\label{eq:growthrateinteracting}
    \check{S}_2 \sim {2\Delta_I\Delta \thinspace t}.
\end{equation}  
Note the contrast with the $\sqrt{t}$ growth in the free model. Eq.~(\ref{eq:growthrateinteracting}) is consistent with linear growth of entanglement entropy expected for generic interacting systems~\cite{kim2013ballistic}.

It would be very interesting to study quantum fluctuations on top of this classical solution, given that it is only unique up to translations.
At the largest length scales, 
we expect that the centre of mass  executes a random walk, whose ``entropic'' fluctuations will correct the growth rate in Eq.~(\ref{eq:growthrateinteracting}) by an amount that is subleading at small $\Delta_I$.\footnote{It would also be interesting to compute the action cost for a domain wall whose centre of mass travels at a non-zero speed: 
this will give the entanglement membrane tension $\mathcal{E}(v)$ as a function of $v$, as well as the butterfly velocity $v_B$.}

\section{Outlook} \label{sec:outlook}

In contrast with the case of interacting systems, which have a discrete replica symmetry, free-fermion systems have a continuous replica symmetry. This leads to smooth domain walls which relax over time, as opposed to the sharp domain walls which give rise to the ``entanglement membrane'' picture in interacting systems.
    
Despite the original problem of calculating entanglement growth being purely quantum mechanical, the equivalent calculation for the Heisenberg chain can be tackled using a semi-classical approximation.
The only large parameter required to justify the semiclassical limit is the large timescale for the evolution.

An interesting feature of the effective theory for free-fermion systems is that we must describe not one classical configuration of our Heisenberg chain, but two: one relaxing forwards in time, and the other relaxing backwards.
    
If we introduce interactions into our model by adding quartic terms into the Hamiltonian, the continuous ${\rm SO}(2N)$ replica symmetry is explicitly broken down to a discrete symmetry. At sufficiently short times $t\ll t_\mathrm{crossover}$, the sharp domain wall created by the boundary between the subsystems $A$ and $B$ relaxes as in the free-fermion case, but over longer times it approaches a finite thickness $l_\mathrm{int}$ which depends on the strength of the interactions. Below this length scale, the domain wall appears smooth, but above this length scale, the domain wall resembles a sharp domain wall similar to those found in strongly interacting systems, indicating a crossover from free-fermion behaviour to interacting behaviour.

This gives for the first time an explicit picture of the renormalization group flow between two different universality classes for the entanglement dynamics --- one associated with Gaussian systems and one associated with interacting ones.

Although we have focussed on the case $N=2$ in order to calculate the averaged purity, a similar qualitative picture should hold for a general number $N$ of replicas of the density matrix,
allowing calculation of, for example, the von Neumann entropy (the limit ${N\to 1}$ of the R\'enyi entropy). 
In principle, extending the present approach would require a generalization of the $\mathrm{SU}(2)$ coherent states parameterization that proved useful above.
    
Interestingly, though, numerics (see Fig.~\ref{fig:fluctuations}) suggest that the average $\check{S}_2$ that we have computed is also equal at leading order to the standard mean $\overline{S_2}$.
This suggests that certain calculations for a larger number of replicas 
in fact reduce 
(at the saddle-point level) 
to the calculation presented here. 
It will be interesting to study how this happens. A natural guess is that the saddle-point solutions needed for the higher moments of the purity in  Eq.~\ref{eq:highermoments} can be ``factorized'' into $k$ copies of the saddle point solution in Sec.~\ref{eq:saddlepointapplications} (see also \cite{sigmamodelmeasurement}). 
If the saddle-point action is simply proportional to $k$, then the generating function (\ref{eq:highermoments}) becomes the generating function of a deterministic variable,
implying that 
fluctuations in $S_2(t)$ --- arising from the randomness in the couplings $\eta(t)$ ---  are  strongly subleading compared to the mean.
It would be interesting to interpret these subleading fluctuations in the effective theory.

The effective theory could be used to compute many other quantities:
 it remains to be seen which quantities are most naturally 
understood via  the  spacetime picture developed here (i.e. at the level of the action) 
and which through other tools (e.g. integrability \cite{bernard2022dynamics}).
In the number-conserving case \cite{Bernard2019Open}, the probability distribution of ``quantum coherences'' $\langle c^\dag_i c_j\rangle$  has non-trivial combinatorial structure
\cite{bernard2021solution,
bernard2021entanglement,
bernard2022dynamics,
hruza2022dynamics,bauer2022bernoulli}.
In the Majorana model, the variance
${\overline{\langle i \gamma_i \gamma_j\rangle^2}}$ maps to an overlap similar to the one we have considered, but with a different boundary state with two flipped spins:
${{\langle{\ldots\uparrow \downarrow_i \uparrow \ldots \uparrow \downarrow_j \uparrow} | e^{-Ht} | {\tilde \psi^+}\rangle}}$. The out-of-time order correlation function could similarly be  computed.

The effective theory we have used throughout this paper seems very different from the quasiparticle picture which has been used successfully to describe entanglement spreading in (non-random) conformal field theories and lattice fermion systems~\cite{calabrese2005evolution,
fagotti2008evolution,
calabrese2009entanglement,
alba2018entanglement}.
One possible way of seeing a connection between these two pictures is to use the fact that the Heisenberg Hamiltonian can also be viewed as the transition operator for a classical Markov process.
This Markov process is the classical Symmetric Simple Exclusion Process \cite{tailleur2008mapping}, 
except that we must retain some ``pairing'' information about the particles as discussed below.
We write the effective ${\rm SO}(3)$ Hamiltonian $H$ in terms of the operator $\varsigma_{i,j}$ which swaps the Heisenberg spins on sites $i$ and $j$,
\begin{equation}
    H = 4\Delta^2 \sum_i [ 1 - \varsigma_{i,i+1}].
\end{equation}
We can then reinterpret the spin wavefunction $e^{-Ht}\vert\tilde{\psi}\rangle$ as a probability distribution,\footnote{If we write the wavefunction in terms of simultaneous eigenstates of $\sigma_i^z$, then the coefficients 
(not their absolute squares)
can be chosen to be real and positive and can be interpreted as probabilities.}
where $H$ randomly swaps neighbouring spins at a rate of $4\Delta^2$ while conserving total probability. 
The initial state $\vert\tilde{\psi}\rangle$,
 which lives in the $N=2$ replicated Hilbert space,
consists of entangled pairs (not pairs of physical Majoranas, but Bell pairs of spin-$\frac{1}{2}$s).
Time evolution under $H$ causes the members of these pairs to hop randomly to the left or right while preserving the entanglement between the two members of the pair. 

Contracting with $\langle C_A \vert$ at the final time 
gives a factor of $1$ for every pair in which both members end up in $A$ or both end up in $B$, 
but gives a factor of $1/\sqrt{2}$ for every pair which has one member in $A$ and one in $B$. 
We find
\begin{equation}
  \overline{e^{-S_2}} = \langle C_A \vert e^{-Ht} \vert \tilde{\psi} \rangle
    = \langle e^{-\frac{1}{2} n \ln 2} \rangle
\end{equation}  
where $n$ is the number of pairs which end up with one member in $A$ and one member in $B$, and where 
the expectation value is taken in this classical Markov process.

It should be borne in mind that this not a physical Markov process for the Majoranas. 
Despite this, the  expression above coincides with the average of $e^{-S_2}$ that would be obtained for a physical Markov process in which we swapped Majoranas. That is a special feature of the average of $\overline{e^{-S_2}}$, and does not hold for the average of $\overline{S_2}$. 
Nevertheless, we see that for $N=2$ there is a formal connection with a particle dynamics that resembles a diffusive version of the quasiparticle picture \cite{calabrese2005evolution,
fagotti2008evolution,
calabrese2009entanglement,
alba2018entanglement}.

The problem with the expression above is that the average on the right hand side is strongly biased towards small values of $n$. Therefore while  $\langle n \rangle$ is easy to compute in the Markov process by standard diffusion arguments, this is not very useful.
The semiclassical picture in previous sections is more useful, partly because it makes explicit the ${\rm SU}(2)$ symmetry that is hidden in the Markov representation.

Let us briefly mention some ways in which the effective theory could be extended. Non-unitarity due to measurement of Majorana bilinears 
\cite{bao2021symmetry, nahum2020entanglement,merritt2023entanglement} in this model will be addressed separately in Ref.~\cite{sigmamodelmeasurement}, where nonlinear sigma model effective field theories will be provided for that case. (In contrast to the theory discussed here, those nonlinear sigma models have a relativistic dispersion.)

Most straightforwardly, the mappings of previous sections also carry over directly to a hypercubic lattice of Majoranas in higher dimensions, and could be used to obtain scaling forms for entanglement of regions of different shapes.

Four-fermion interactions have been discussed above, 
but a different way to break the Gaussian structure 
is to explicitly break the $\mathbb{Z}_2$ symmetry of the Ising model (recall that the original Majorana chain is dual to a noisy Ising model).
 This is a non-local perturbation in the Majorana language. It would be interesting to explore how the number of effective ground states is reduced (Sec.~\ref{sec:symmetriesgeneral}). 
We believe that this process can be understood as a confinement phenomenon in the replica theory 
(we will discuss this elsewhere).

Retaining a non-random hopping in addition to the random one we have considered would mean that, at least at short scales, it was necessary to retain fermionic degrees of freedom in the effective theory; it would be interesting to study the reduction to a bosonic effective theory at large scales.
It would also be interesting to study the case of static disorder
(on lengthscales where Anderson localization effects are weak or in dimensionalities where  localization is avoided). 
With standard kinds of approximations \cite{lerner2003nonlinear}, this leads to an effective model where the interactions are long-range in time. However, the replica symmetry is still continuous, and the saddle-point approach developed here could be extended to this case.

\medskip

{\bf Acknowledgments:} 
We are extremely grateful to John Chalker for valuable discussions throughout this project, and to Fabian Essler for extremely useful discussions that helped us clarify the coherent states path integral.
TS was supported by a James Buckee Scholarship and by Royal Society Enhancement Award.
AN was supported in part by a Royal Society University Research Fellowship and by CNRS and the ENS. 
DB was in part supported by CNRS, by the ENS, and by the ANR project “ESQuisses”, contract number ANR-20-CE47-0014-01.

\appendix
\addappheadtotoc

\section{Stabilizers for $\vert C_A \rangle$ in terms of Majorana operators} \label{appindexcontractionmajoranas}

We wish to  find more explicit expressions for the bras corresponding to index contractions on the replicas. We do this by determining the stabilizers of such states  in terms of Majorana operators.

It will be convenient to use the following mapping from the original Majorana chain to an Ising spin chain:
\begin{align}
    i\gamma_{2j-1}\gamma_{2j} &= Z_j \label{sigmaspinz}\\
    i\gamma_{2j}\gamma_{2j+1} &= X_j X_{j+1} \label{sigmaspinxx}
\end{align}
where $X_j$ and $Z_j$ are Pauli matrices. The Pauli matrices act on spins $j$ formed from pairs of Majorana modes.

When we create $2N$ replicas of the chain and carry out the sign change $\gamma_{2j}^{2n} \to \gamma_{2j}^{2n}$ this becomes:
\begin{align}
    i\gamma_{2j-1}^a\gamma_{2j}^a &= (-1)^{a+1} Z_j^a \label{sigmaspinsz} \\
    i\gamma_{2j}^a\gamma_{2j+1}^a &= (-1)^{a+1} X_j^a X_{j+1}^a \label{sigmaspinsxx}
\end{align}
The sign change effectively changes the order of the gammas on even replicas with respect to the mapping to the Ising chain.

How do we write index contractions as bras? Take the example of a single spin with $N=1$ (so there are two replicas). How do we write the identity operator as a state $\langle \mathbb{1} \vert$?

We insist that $\langle \mathbb{1} \vert(\vert \psi \rangle \otimes \vert \phi^* \rangle) = \langle \phi \vert \psi \rangle$ where $\vert\psi\rangle$ and $\vert\phi\rangle$ are arbitrary spins. This means that $\vert \mathbb{1} \rangle = \mid\uparrow\uparrow\rangle + \mid\downarrow\downarrow\rangle$ (where the left and right spins refer to replicas 1 and 2 respectively).

The state $\vert\mathbb{1}\rangle$ can be uniquely defined by its stabilizers
\begin{align}
    Z^1 Z^2 \vert\mathbb{1}\rangle &= \vert\mathbb{1}\rangle \\
    X^1 X^2 \vert\mathbb{1}\rangle &= \vert\mathbb{1}\rangle
\end{align}
where $X^a$ and $Z^a$ are once again Pauli matrices acting on replica $a$. For a chain of Majoranas with 2 replicas, we similarly get a pair of stabilizers for every pair of Majorana sites, assuming we are applying the identity contraction everywhere.

The stabilizer $Z_j^1 Z_j^2$ can be straightforwardly written in terms of Majorana operators using equation (\ref{sigmaspinsz}) to give $\gamma_{2j-1}^1\gamma_{2j}^1\gamma_{2j-1}^2\gamma_{2j}^2$.

The stabilizer $X_j^1 X_j^2$ is not as simple. We can write products of different $X$ matrices on the same replica by stringing together products of neighbouring $X$ operators e.g. $X_{j}X_{j+3} = (X_{j}X_{j+1})(X_{j+1}X_{j+2})(X_{j+2}X_{j+3})$ and then use equation (\ref{sigmaspinsxx}). However, we cannot do this for products of $X$ operators in different chains because there are no relations of the form (\ref{sigmaspinsxx}) which relate Majoranas from different replicas.

We can fix this by ``connecting'' the two chains on the left-hand side by defining $i\gamma_1^2\gamma_1^1=X_1^1 X_1^2$. This essentially means we use the mapping in equations (\ref{sigmaspinz}) and (\ref{sigmaspinxx}) but treating both replicas as part of a single chain.

We can now iteratively find a simple set of stabilizers $A_i \equiv -i\gamma_i^1\gamma_i^2$ in terms of Majorana operators:
\begin{itemize}
    \item $A_1$ is a stabilizer because $A_1 = -i\gamma_1^1\gamma_1^2 = X_1^1 X_1^2$
    \item If $A_i$ is a stabilizer, then $A_{i+1}$ is also a stabilizer.
\end{itemize}

The second point follows from the relations
\begin{align}
    A_{2j-1} &= (X_{j-1}^1 X_{j-1}^2) (X_j^1 X_j^2) A_{2j-2} \\
    A_{2j} &= (Z_j^1 Z_j^2) A_{2j-1}
\end{align}
which follow from equations (\ref{sigmaspinsz}) and (\ref{sigmaspinsxx}) and the definition of $A_i$.

This logic extends to a larger number of replicas.
Let us consider the state
$\ket{C_A}$ that we need for the computation of the purity.
Region $B$ consists of the first $\ell$ Ising sites (or the first $2\ell$ Majorana sites) and region $A$ consists of the remainder. 
In the Ising language, the state at a given site is as follows. In $B$, we form a maximally entangled state of the form $\ket{\mathds{1}}$ between layers $a=1$ and $a=2$, and separately also between layers $a=3$ and $a=4$. In $A$, we similarly entangle $a=1$ with $a=4$ and $a=2$ with $a=3$ (Fig.~\ref{figindexcontractions}).
As a result, the stabilizers are
\ba\label{eq:isingstabilizersleft}
& Z^1Z^2,& &X^1 X^2, & 
& Z^3Z^4,& &X^3 X^4 & &(\text{region $B$}), 
\\\label{eq:isingstabilizersright}
& Z^2Z^3,& &X^2 X^3, & 
& Z^1Z^4,& &X^1 X^4 & &(\text{region $A$}). 
\end{align}
In analogy to the above, we connect the chains at both the left and the boundaries by defining
\ba\label{eq:isingboundaryops}
X_1^1 X_1^2 & = i \gamma_1^2 \gamma_1^1,
& 
X_1^3 X_1^4 & = i \gamma_1^4 \gamma_1^3,
& 
X_L^2 X_L^3 & = i \gamma_{2L}^3 \gamma_{2L}^2.
\end{align}
Then we can show that in the Majorana language we have the stabilizers (with eigenvalue $+1$)
\ba
& i \gamma_i^2 \gamma_i^1, 
& 
& \gamma_i^4 \gamma_i^3
& &(i\in B),\\
& i \gamma_i^3 \gamma_i^2, 
& 
& \gamma_i^4 \gamma_i^1
& &(i\in A)
\end{align}
at a site $i$.
These can be determined in the same way as for $N=1$. 
For the stabilizers in region $B$ we  start at $i=1$  (where the above equations hold by Eq.~(\ref{eq:isingboundaryops})) and propagate to the right by multiplying with the stabilizers in Eq.~(\ref{eq:isingstabilizersleft}).
For the stabilizers of the form $i\gamma^3\gamma^2$ in region $A$ we start at the right and propagate to the left similarly. 
The stabilizers of the form 
$i\gamma^4_i\gamma^1_i$ in region $A$ we write as a snake-like product of all the Majoranas in the 4 layers between site $2\ell$ and site $i$ inclusive (note that ${i>2\ell}$ since $i$ is in region $A$). Then we may check that the terms in this product all cancel by using either the stabilizers in Eqs.~(\ref{eq:isingstabilizersleft}),~(\ref{eq:isingstabilizersright}), or by using the already-known stabilizers $i\gamma^3_i\gamma^2_i$, 
$i\gamma^2_{2\ell}\gamma^1_{2\ell}$, $i\gamma^4_{2\ell}\gamma^3_{2\ell}$.

\section{Expectation values of operators as Ising configurations} \label{appexpectationvaluesising}

For $N=1$ we get an Ising spin chain with spin operators $A_i \equiv -i\gamma_i^1\gamma_i^2$. We can work in the basis of simultaneous eigenstates of these operators.

We can write any operator acting on the original Majorana chain as a linear combination of products of $\gamma$ operators, so we can choose our basis of operators to be the set of products of the $\gamma$ operators. Given that $(\gamma_i)^2=1$, there are $2^L$ such operators (including the identity operator).

How can we write these operators as bras in the replicated system? Given that $\langle \phi \vert \hat{O} \vert \psi \rangle = \langle \phi \vert  \mathbb{1} (\hat{O} \vert \psi \rangle)$, if we know how to write the identity operator $\mathbb{1}$ as a bra $\langle\mathbb{1}\vert$, then we can generate the bra for any other operator $\hat{O}$ using $\langle \hat{O} \vert = \langle \mathbb{1} \vert \hat{O}^1$ where the operator $\hat{O}^1$ acts only on the first replica, and the bras are now states in the doubled representation.

As explained in the previous appendix, the identity state $\vert\mathbb{1}\rangle$ satisfies $A_i\vert\mathbb{1}\rangle = +\vert\mathbb{1}\rangle$ for all $i$. In other words, the Ising spin on every site $i$ is in the spin up state.

Let us now consider how acting with Majorana operators affects arbitrary Ising configurations. Take some Ising configuration $\vert\{s_i\}\rangle$ where $A_i\vert\{s_i\}\rangle=s_i\vert\{s_i\}\rangle$. Now consider the state $\gamma_j^1\vert\{s_i\}\rangle$. The new values of $s'_i$ can be determined from
\begin{align*}
    A_i(\gamma_j^1\vert\{s_i\}\rangle)&=(-1)^{\delta_{ij}}\gamma_j^1 A_i\vert\{s_i\}\rangle \\
    &=(-1)^{\delta_{ij}} \gamma_j^1 s_i\vert\{s_i\}\rangle \\
    &=s'_i(\gamma_j^1\vert\{s_i\}\rangle)
\end{align*}
so $s'_i=(-1)^{\delta_{ij}}s_i$. This means that acting with $\gamma_j^1$ simply flips the Ising spin at $j$. (This doesn't tell us anything about the overall phase of the state. Acting with $\gamma_j^1$ then $\gamma_k^1$ gives the same state as acting with $\gamma_k^1$ then $\gamma_j^1$ but with the opposite sign.)

If we act on the state $\vert\mathbb{1}\rangle$ (all up spins) with some arbitrary product of Majorana operators $\hat{O}$, the resulting state $\vert\hat{O}\rangle$ is therefore a single Ising configuration with $s_i=-1$ if $\gamma_i$ appears in the this product, and $s_i=+1$ otherwise. This tells us how to convert from products of Majorana operators to states in the replicated system, up to complex number prefactors which drop out of the final result. To see this, write a general operator $\hat O$ of interest as a sum ${\hat O = \sum_\alpha c_\alpha \hat O_\alpha}$ of  products of Majoranas. After the Ising mapping, each $\langle{\hat O_\alpha}|$ has a definite Ising energy $E_\alpha$, whose computation is explained in Sec.~\ref{sec:expectationvalues}, so that the physical expectation value at time $t$ is
\be
\langle \hat O\rangle (t) = \sum_{\alpha} c_\alpha e^{-E_\alpha t} \langle \hat O_\alpha\rangle(0).
\ee

\section{Reduction from an ${\rm SO}(4)$ spin to two ${\rm SU}(2)$ spins} \label{app:symmetrygroupreduction}

For $N=2$, we have six operators $A^{ab}$ with $a,b=1,\dots,4$, such that $J^{ab}\equiv\frac{1}{2}A^{ab}$ satisfy the angular momentum commutation relations (\ref{eq:amcommutationrelations}) and therefore form a representation of ${\rm so}(4)$.

We can define a set of Pauli operators $\sigma^i$ which act on spins with $\chi=+1$
\begin{align}
    \sigma^x = \frac{1}{2} (A^{23} + A^{14}) \\
    \sigma^y = \frac{1}{2} (A^{31} + A^{24}) \\
    \sigma^z = \frac{1}{2} (A^{12} + A^{34})
\end{align}
These operators commute with $\chi$ and satisfy $\sigma^i\frac{1-\chi}{2}=0$ (so they annihilate states with $\chi=-1$). They also satisfy $[\sigma^i,\sigma^j]=2i\epsilon_{ijk}\sigma^k$ and $(\sigma^i)^2=\frac{1+\chi}{2}$ so act as Pauli operators on the subspace of states with $\chi=+1$.

Similarly we can define a set of Pauli operators $\tilde\sigma^i$ which act on spins with $\chi=-1$
\begin{align}
    \tilde\sigma^x = \frac{1}{2} (A^{23} - A^{14}) \\
    \tilde\sigma^y = \frac{1}{2} (A^{31} - A^{24}) \\
    \tilde\sigma^z = \frac{1}{2} (A^{12} - A^{34})
\end{align}
which again commute with $\chi$ but satisfy $\sigma^i\frac{1+\chi}{2}=0$ (so they annihilate states with $\chi=+1$). They satisfy $[\tilde\sigma^i,\tilde\sigma^j]=2i\epsilon_{ijk}\tilde\sigma_-^k$ but $(\tilde\sigma^i)^2=\frac{1-\chi}{2}$ so act as Pauli operators on the subspace of states with $\chi=-1$.

We have reduced the ${\rm SO}(4)$ spin down to a superposition of two conventional ${\rm SU}(2)$ spin-$\frac{1}{2}$s. With $\sigma^i$ acting on the first type and $\tilde\sigma^i$ on the second type. The Heisenberg interaction term in (\ref{generalhamiltonian}) becomes
\begin{equation}
    \sum_{a<b}^{2N}A_i^{ab}A_{i+1}^{ab} = 2\left[ \vec{\sigma}_i\cdot\vec{\sigma}_{i+1} + \vec{\tilde{\sigma}}_i\cdot\vec{\tilde{\sigma}}_{i+1} \right]
\end{equation}
so two neighbouring spins with the same chirality simply experience a Heisenberg interaction, while neighbouring spins of opposite chirality do not interact.

\section{The state $\vert C_A \rangle$ in the ${\rm SO}(4)$ Heisenberg model} \label{appindexcontractiondomainwall}

Let us consider a single Majorana site.

As explained in Appendix \ref{appindexcontractionmajoranas}, the state $\vert \mathbb{1} \rangle$ which acts like the identity at a given Majorana site can be defined by
\begin{align}
    -i\gamma^1\gamma^2\vert \mathbb{1} \rangle &= +\vert \mathbb{1} \rangle \\
    -i\gamma^3\gamma^4\vert \mathbb{1} \rangle &= +\vert \mathbb{1} \rangle
\end{align}
Clearly this state satisfies $\chi\vert \mathbb{1} \rangle = -\gamma^1\gamma^2\gamma^3\gamma^4\vert \mathbb{1} \rangle=+\vert \mathbb{1} \rangle$.

Similarly, the state $\vert C \rangle$ which swaps replicas at a given Majorana site can be defined by
\begin{align}
    -i\gamma^2\gamma^3\vert C \rangle &= +\vert C \rangle \\
    -i\gamma^1\gamma^4\vert C \rangle &= +\vert C \rangle
\end{align}
which clearly also satisfies $\chi \vert C \rangle = + \vert C \rangle$.

Using the Pauli operators defined in Appendix \ref{app:symmetrygroupreduction}, the fact that $\sigma^z\vert\mathbb{1}\rangle=+\vert\mathbb{1}\rangle$ and $\sigma^x\vert C\rangle=+\vert C\rangle$ allows us define
\begin{align}
    \vert\mathbb{1}\rangle &= \mid\uparrow\rangle \\
    \vert C\rangle &= \mid\rightarrow\rangle
\end{align}
In other words, the state $\vert\mathbb{1}\rangle$ corresponds to an ``up'' spin, and the state $\vert C\rangle$ corresponds to a ``right'' spin.

The state $\vert C_A\rangle$ is a product state of $\vert C\rangle$ on every site inside the subsystem $A$ and $\vert\mathbb{1}\rangle$ on every site outside the subsystem $A$, so in the Heisenberg model it is similarly a product state of $\mid\rightarrow\rangle$ on every site inside $A$ and $\mid\uparrow\rangle$ on every site outside $A$.

\section{The state $\vert \tilde{\psi} \rangle$ in the Heisenberg model} \label{appreplicatedinitialstate}

We wish to find the state $\vert \tilde{\psi}^+ \rangle$, that is, the replicated initial state $\vert \tilde{\psi} \rangle$ projected onto the subspace where all spins have positive chirality $\chi_i=+1$. To do this, we will first consider a single pair of spins.

Given that the initial state $\vert\psi\rangle$ is defined by $i\gamma_1\gamma_2\vert\psi\rangle = +\vert\psi\rangle$, the replicated initial state $\vert\tilde{\psi}\rangle$ can be defined (up to a phase) by the stabilizers
\begin{align}
    i\gamma_1^1\gamma_2^1\vert\tilde{\psi}\rangle &= +\vert\tilde{\psi}\rangle \\
    -i\gamma_1^2\gamma_2^2\vert\tilde{\psi}\rangle &= +\vert\tilde{\psi}\rangle \\
    i\gamma_1^3\gamma_2^3\vert\tilde{\psi}\rangle &= +\vert\tilde{\psi}\rangle \\
    -i\gamma_1^4\gamma_2^4\vert\tilde{\psi}\rangle &= +\vert\tilde{\psi}\rangle
\end{align}
where the minus signs are a result of the redefinition $\gamma_{2j}^{2n} \mapsto -\gamma_{2j}^{2n}$.

This immediately tells us that $\chi_1\chi_2\vert\tilde{\psi}\rangle=+\vert\tilde{\psi}\rangle$, so the chiralities of the two spins are perfectly correlated.

Using the Pauli matrices $\sigma_1^i$ and $\sigma_2^i$ which act only on the positive chirality parts of the spins (as defined in Appendix \ref{app:symmetrygroupreduction}) we can write
\begin{align}
    \sigma_1^x\sigma_2^x &= \frac{1+\chi_1}{4}\left[(-i\gamma_1^2\gamma_2^2)(i\gamma_1^3\gamma_2^3)+(i\gamma_1^1\gamma_2^1)(-i\gamma_1^4\gamma_2^4)\right] \\
    \sigma_1^z\sigma_2^z &= \frac{1+\chi_1}{4}\left[(-i\gamma_1^1\gamma_2^1)(i\gamma_1^2\gamma_2^2)+(i\gamma_1^3\gamma_2^3)(-i\gamma_1^4\gamma_2^4)\right]
\end{align}
acting with both sides on $\vert\tilde{\psi}\rangle$ gives
\begin{align}
    \sigma_1^x\sigma_2^x\vert\tilde{\psi}\rangle &= \frac{1+\chi_1}{2}\vert\tilde{\psi}\rangle \\
    \sigma_1^z\sigma_2^z\vert\tilde{\psi}\rangle &= \frac{1+\chi_1}{2}\vert\tilde{\psi}\rangle
\end{align}
But the fact that $\chi_1\chi_2\vert\tilde{\psi}\rangle=+\vert\tilde{\psi}\rangle$ means that projecting onto $\chi_1=+1$ also projects onto $\chi_2=+1$ so
\begin{equation}
    \frac{1+\chi_1}{2}\vert\tilde{\psi}\rangle = \vert\tilde{\psi}^+\rangle
\end{equation}
which means
\begin{align}
    \sigma_1^x\sigma_2^x\vert\tilde{\psi}^+\rangle &= +\vert\tilde{\psi}^+\rangle \\
    \sigma_1^z\sigma_2^z\vert\tilde{\psi}^+\rangle &= +\vert\tilde{\psi}^+\rangle
\end{align}
These stabilizers determine $\vert\tilde{\psi}^+\rangle$ up to a phase and normalization. The fact that quantum states must have norm 1 means that $\langle\tilde{\psi}^+\vert\mathbb{1}\rangle=1$ (where $\vert\mathbb{1}\rangle=\mid\uparrow\uparrow\rangle$) which fixes the state to be
\begin{equation}
    \vert\tilde{\psi}^+\rangle = \mid\uparrow\uparrow\rangle\ +\mid\downarrow\downarrow\rangle
\end{equation}

It is worth noting that if we had chosen the orthogonal state $i\gamma_1\gamma_2\vert \psi\rangle = -\vert \psi\rangle$ we would still get the same $\vert\tilde{\psi}^+\rangle$. This implies that $\vert\tilde{\psi}^+\rangle$ is not sensitive to the specific initial state $\vert \psi\rangle$ we choose, but rather to the entanglement of this state.

For the general case of $L/2$ pairs, the state $\vert\tilde{\psi}^+\rangle$ is simply a product of the above for each pair.

\section{Equilibrium purity including fluctuations} \label{apppurityincludingfluctuations}

We wish to calculate $\langle \tilde{\psi}^+ \vert e^{-Ht} \vert C_A \rangle$ in the limit $t \to \infty$. Here $\vert C_A \rangle$ is the sharp domain wall state and $\vert \tilde{\psi}^+ \rangle$ is a product of entangled pairs. We can write this as
$\langle \tilde{\psi}^+ \vert \hat{P} \vert C_A \rangle$ where $\hat{P}$ is the projector onto ground states of the Heisenberg chain.

Instead of using spin coherent states with spin $j=\frac{1}{2}$, we can write $\hat{P}$ as an integral over coherent states of spin $j=L/2$, where $L$ is the number of sites:
\begin{equation}
    \langle \tilde{\psi}^+ \vert \hat{P} \vert C_A \rangle = \frac{2j+1}{\pi}\int\frac{d^2 z}{(1+\bar{z}z)^{2j+2}} \langle \tilde{\psi}^+ \vert z \rangle \langle z \vert C_A \rangle
\end{equation}
So we need to find the overlaps $\langle \tilde{\psi}^+ \vert z \rangle$ and $\langle z \vert C_A \rangle$.

To calculate $\langle z \vert C_A \rangle$, take the state $\vert C_A \rangle$ to have $z=-c$ in the left side of the chain and $z=+c$ in the right side of the chain, where $c=\sqrt{2}-1$.
This gives an overlap
\begin{align}
    \langle z \vert C_A \rangle &=
    \left ( \frac{1}{\sqrt{1+c^2}}\langle z \vert -c \rangle \right )^{L/2}
    \left ( \frac{1}{\sqrt{1+c^2}}\langle z \vert +c \rangle \right )^{L/2} \\
    &= (1+c^2)^{-L/2} (1-c\bar{z})^{L/2} (1+c\bar{z})^{L/2} \\
    & = (1+c^2)^{-L/2} (1-c^2\bar{z}^2)^{L/2} \\
    & = (1+c^2)^{-j} (1-c^2\bar{z}^2)^{j}
\end{align}

To calculate $\langle \tilde{\psi}^+ \vert z \rangle$, take the state $\vert \tilde{\psi}^+ \rangle$ to be a product of pairs in the state $\vert \rightarrow\rightarrow \rangle + \vert \leftarrow\leftarrow \rangle$. In terms of single-spin coherent states this is $\frac{1}{2}\vert +1 \rangle \vert +1 \rangle + \frac{1}{2}\vert -1 \rangle \vert -1 \rangle$. Therefore we get
\begin{align}
    \langle \tilde{\psi}^+ \vert z \rangle &=
    \left (
    \frac{1}{2} \langle +1 \vert z \rangle^2
    + \frac{1}{2} \langle -1 \vert z \rangle^2
    \right )^{L/2} \\
    &= 
    \left (
    \frac{1}{2} (1+z)^2 + \frac{1}{2} (1-z)^2
    \right )^{L/2} \\
    & = (1+z^2)^{L/2} \\
    & = (1+z^2)^{j}
\end{align}

This gives us the exact expression
\begin{equation}
\begin{split}
    \langle \tilde{\psi}^+ \vert \hat{P} \vert C_A \rangle &=\frac{2j+1}{\pi}(1+c^2)^{-j} \\
    \times \int&\frac{d^2 z}{(1+\bar{z}z)^{2j+2}} (1+z^2)^{j} (1-c^2\bar{z}^2)^{j}
\end{split}
\end{equation}

To study the thermodynamic limit we take $j \to \infty$. But before we can do this we need to change variables to
\begin{align}
    \zeta &= z \sqrt{j+1/2}\\ \bar{\zeta} &=  \bar{z} \sqrt{j+1/2}
\end{align}
If we do this and use $(1+x/n)^n \approx e^x$ for large $n$ we get
\begin{equation}
\begin{split}
    \langle \tilde{\psi}^+ \vert \hat{P} \vert C_A \rangle \approx \frac{2}{\pi}(1+c^2)^{-j}\int d^2 \zeta
    e^{-2\bar{\zeta}\zeta+\zeta^2-c^2\bar{\zeta}^2} \\
    \times \left(1+\frac{1}{j+1/2}\zeta^2\right)^{-1/2}
    \left(1-\frac{c^2}{j+1/2}\bar{\zeta}^2\right)^{-1/2}
\end{split}
\end{equation}

For large $j$ the bracketed terms on the right are close to one for typical values of $\zeta$ and we can approximate the integral as a basic Gaussian integral
\begin{equation}
    \langle \tilde{\psi}^+ \vert \hat{P} \vert C_A \rangle \approx \frac{2}{\pi}(1+c^2)^{-j}\int d^2 \zeta
    e^{-2\bar{\zeta}\zeta+\zeta^2-c^2\bar{\zeta}^2}
\end{equation}

Rewriting $\zeta=x_1+ix_2$ we get
\begin{equation}
    \langle \tilde{\psi}^+ \vert \hat{P} \vert C_A \rangle \approx \frac{2}{\pi}(1+c^2)^{-j}\int d^2 x
    e^{-\mathbf{x}^{\mathrm{T}}\mathbf{M}\mathbf{x}}
\end{equation}
where
\begin{equation}
    \mathbf{M} =
    \begin{pmatrix}
        1+c^2 & -i(1+c^2) \\
        -i(1+c^2) & 3-c^2
    \end{pmatrix}
\end{equation}
and we get $\mathrm{det}(\mathbf{M}) = 4(1+c^2)$. Performing the integral we get
\begin{align}
    \langle \tilde{\psi}^+ \vert \hat{P} \vert C_A \rangle &\approx \frac{2}{\pi}(1+c^2)^{-j} \cdot
    \sqrt{\frac{\pi^2}{4(1+c^2)}} \\
    &= (1+c^2)^{-(j+1/2)} \\
    &= (1+c^2)^{-\frac{L+1}{2}}
\end{align}
so the log of the averaged purity is proportional to $L+1$ as $L$ becomes large.

\section{The $\tau=t$ boundary condition} \label{appfinaltimeboundary}

Let's first consider a single pair of spins. The state $\langle \tilde{\psi}^+\vert$ is then simply
\begin{equation}
    \langle \tilde{\psi}^+\vert = \langle\uparrow\uparrow\mid + \langle\downarrow\downarrow\mid
\end{equation}
Taking the $z$ values of the spins to be $w_1$ and $w_2$ respectively, the un-normalized state is
\begin{equation}
    \vert w_\mathrm{un-normalized}\rangle = w_1 w_2 \mid\uparrow\uparrow\rangle + w_1 \mid\uparrow\downarrow\rangle
    + w_2 \mid\downarrow\uparrow\rangle + \mid\downarrow\downarrow\rangle
\end{equation}
which means that
\begin{equation}
    f(w_1,w_2) = \ln (1 + w_1 w_2)
\end{equation}
Substituting this into eq. (\ref{generaltauequalstboundary}) gives
\begin{align}
    \frac{w_2}{1+w_1 w_2} &= \frac{\bar{w}_1}{1+\bar{w}_1 w_1} \\
    \frac{w_1}{1+w_1 w_2} &= \frac{\bar{w}_2}{1+\bar{w}_2 w_2}
\end{align}
which gives $\bar{w}_1 = w_2$ and $\bar{w}_2 = w_1$. So the boundary condition simply ``swaps'' $w_1$ and $w_2$ to give $\bar{w}_1$ and $\bar{w}_2$.

In general, when we have $L/2$ pairs of spins, the state  $\langle \tilde{\psi}^+\vert$ is simply a product state of entangled pairs, each in the state $\langle\uparrow\uparrow\mid + \langle\downarrow\downarrow\mid$. The function $f$ is simply a sum of the $f$s for each pair of spins, so the above argument applies to each pair of spins individually, and we get $\bar{w}_{2j-1}=w_{2j}$ and $\bar{w}_{2j}=w_{2j-1}$.

\section{Scaling function $g(\alpha)$ and numerical approximation of $\kappa$} \label{app:scalingfunction}

\begin{figure}
\centering
\includegraphics[width=\columnwidth]{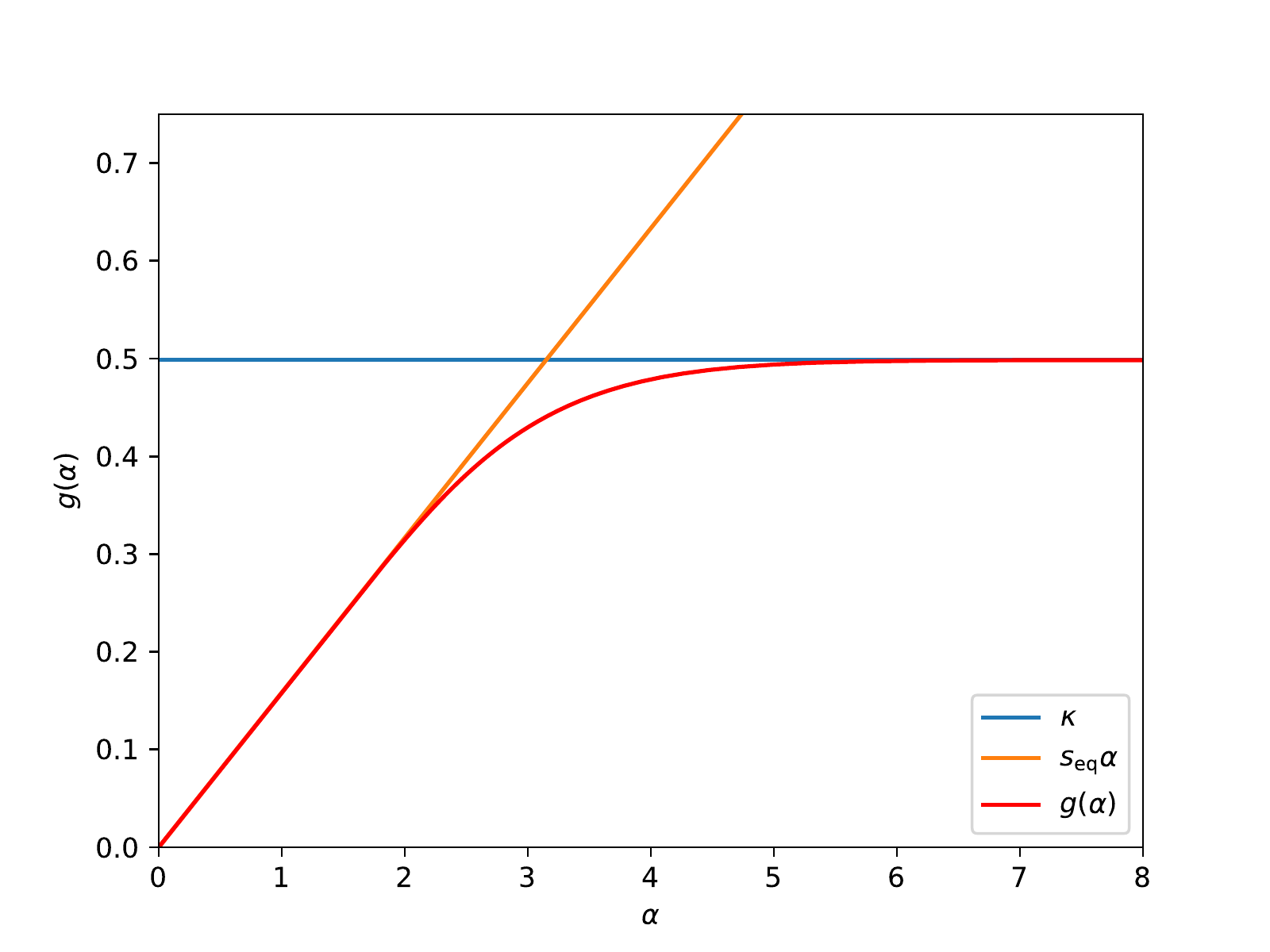}
\caption{Plot of the scaling function $g(\alpha)$. As $\alpha\to\infty$, $g(\alpha)$ approaches a constant value $\kappa$, corresponding to behaviour in a infinite system. As $\alpha\to 0$, $g(\alpha)$ is linear in $\alpha$, given that the entanglement has saturated and is independent of $t$. The slope as $\alpha\to 0$ is half the equilibrium density $s_\mathrm{eq}=\mathcal{S}_\mathrm{eq}/L_A$.}\label{fig:scalingfunc}
\end{figure}

We can numerically approximate the scaling function $g(\alpha)$ by numerically finding $\mathcal{S}$ for a given $L_A$ at different values of $t$, then dividing by $\Delta\sqrt{t}$. As $L_A\to\infty$ this will give the scaling function $g(\alpha)$ exactly. A plot of $g(\alpha)$ using $L_A=50$ is shown in Figure \ref{fig:scalingfunc}.

For each value of $L_A$, we can fit a value of $\kappa_{L_A}$, this being the value that $g_{L_A}(\alpha)$ approaches when $\alpha$ becomes large, so that $\Delta^2 t \ll L^2$ (but not so large that $\Delta^2 t \sim 1$ and the lattice become important).

For fixed ${\alpha\gg 1}$, we find that ${g_{L_A}(\alpha)\approx g(\alpha)-\mathrm{const.}/L^2}$. Taking ${\alpha\gg 1}$ gives ${\kappa\approx 0.49855\dots}$.

\bibliography{noisymajorana.bib}

\end{document}